\documentclass[journal]{IEEEtran}
\ifCLASSINFOpdf
\else
\fi

\usepackage[english]{babel}
\usepackage{amssymb}
\usepackage{amsmath}
\usepackage{mathrsfs}
\usepackage{epsfig}
\usepackage{graphicx}
\usepackage[thinlines]{easytable}
\usepackage{enumitem}
\usepackage{accents}

\newlength{\dhatheight}
\newcommand{\doublehat}[1]{%
    \settoheight{\dhatheight}{\ensuremath{\widehat{#1}}}%
    \addtolength{\dhatheight}{-0.35ex}%
    \widehat{\vphantom{\rule{1pt}{\dhatheight}}%
    \smash{\widehat{#1}}}}

\newtheorem{lemma}{Lemma}

\newtheorem{definition}{Definition}
\newtheorem{theorem}{Theorem}

\def\argmin{\operatornamewithlimits{arg\,min}}

\title{Sign-Perturbed Sums: A New System Identification Approach for Constructing Exact Non-Asymptotic Confidence Regions in Linear Regression Models} 

\author{Bal\'azs Cs. Cs\'aji~\IEEEmembership{Member,~IEEE},\hspace{10mm} \and Marco C. Campi~\IEEEmembership{Fellow,~IEEE},\hspace{10mm} \and Erik Weyer~\IEEEmembership{Member,~IEEE}%
\thanks{The work of B.\ Cs. Cs\'aji was
partially supported by the ARC grants DE120102601 and DP0986162, and the J\'anos Bolyai
Fellowship of the Hungarian Academy of Sciences, BO/00683/12/6. The work of M.\ C.\ Campi was partly supported
by MIUR - Ministero dell'Istruzione, dell'Universit\`a e della Ricerca. The work of E.\ Weyer was supported by the Australian Research Council
(ARC) under Discovery Grants DP0986162 and DP130104028. }%
\thanks{B.\ Cs.\ Cs\'aji is with MTA SZTAKI: The Institute for Computer Science
and Control of the Hungarian Academy of Sciences, Kende utca 13-17,
H-1111 Budapest, Hungary. balazs.csaji@sztaki.mta.hu}%
\thanks{M.\ C.\ Campi is with Department of Information Engineering,
University of Brescia, Via Branze 38, 25123 Brescia, Italy.
marco.campi@unibs.it}%
\thanks{E.\ Weyer is with Department of Electrical and Electronic Engineering, The University of Melbourne, VIC 3010, Australia. ewey@unimelb.edu.au}}%

\begin{document}
\maketitle

\begin{abstract}
We propose a new system identification method, called Sign\,-\,Perturbed Sums (SPS), for constructing  non-asymptotic confidence regions under mild statistical assumptions. SPS is introduced for  linear regression models, including but not limited to FIR systems, and we show that the SPS confidence regions have exact confidence probabilities, i.e., they contain the true parameter with a user-chosen exact probability for any finite data set. Moreover, we also prove that the SPS regions are star convex with the Least-Squares (LS) estimate as a star center. The main assumptions of SPS are that the noise terms are independent and  symmetrically distributed about zero, but they can be nonstationary, and their distributions need not be known. The paper also proposes a computationally efficient ellipsoidal outer approximation algorithm for SPS. 
Finally, SPS is demonstrated through a number of simulation experiments.
\end{abstract}

\begin{IEEEkeywords}
system identiﬁcation, finite sample properties, parameter estimation, linear regression models, least squares methods, statistics.
\end{IEEEkeywords}


\section{Introduction}

\IEEEPARstart{E}{stimating}
parameters of partially unknown systems based on noisy observations is a classical problem in  signal processing, system identification, machine learning and statistics. There are several standard methods available, which typically provide point estimates. Given an estimate, it is an intrinsic task to evaluate how close the estimated parameter is to the true one and such evaluation often comes in the form of confidence regions. Confidence regions are especially important for problems involving strict safety, stability or quality guarantees, and serve as a basis for ensuring robustness.

In practice, we only have a finite number of measurements and limited statistical knowledge about the noise, and this strongly restricts the number of methods  available for constructing confidence regions, unless we are satisfied with approximate, heuristic solutions. Here, we propose a new statistical parameter estimation approach, called {\em Sign-Perturbed Sums} (SPS), for constructing finite-sample, quasi distribution-free confidence regions. This paper introduces and analyzes the SPS method for linear regression models including Finite Impulse Response (FIR) and Generalised FIR models. 

Linear regression is a classical problem in statistics, which finds applications in many fields. It is a core component of identification, learning and prediction tasks, and a typical application is to estimate parameters of dynamical systems from experimental data, which is one of the fundamental problems of system identification \cite{Ljung1999,Ljung1994,Ljung2010,Soderstrom1989,Gevers2006}. Under natural conditions the Least-Squares (LS) method provides a strongly consistent point estimate of the system parameters. Moreover, the parameter estimation error is asymptotically normal, and this property can be used to build approximate confidence regions. However, these regions are
based on the Central Limit Theorem, and hence are guaranteed only asymptotically as the number of  data points tends to infinity. Therefore, applying the classical system identification theory with finitely many data points results only in heuristic confidence regions, which do not come with strict theoretical guarantees. This calls for alternative approaches that allow us to construct
 {\em guaranteed}, {\em non-asymptotic} confidence regions around the Least-Squares Estimate (LSE)
under  {\em mild  statistical assumptions}.

The system identification method ``Leave-out Sign-dominant Correlation Regions'' (LSCR) was developed earlier \cite{Campi2005,Campi2010,Campi2009,Dalai2007} by  authors of this paper. This method can build guaranteed, non-asymptotic confidence regions for parameters of various (linear and non-linear) dynamical systems under weak assumptions on the noise. 
LSCR has been applied successfully in various contexts \cite{Weyer2009,Weyer2009b}, and the topic has been the object of various investigations and related studies \cite{Schoukens2013,Dabbene2014,Aguero2012,Granichin2012,DenDekker2008,Hjalmarsson2006}. However, LSCR provides confidence regions with exact probabilities only for scalar parameters, while it offers bounds for the multidimensional case. 
Furthermore, the inclusion of the LSE in the confidence regions constructed by LSCR is not guaranteed.

The non-asymptotic SPS method, which is presented in this paper,  provides exact confidence regions for multidimensional parameter vectors and guarantees the inclusion of the LSE in the confidence set. The exact finite-sample confidence probability is guaranteed even though the knowledge of the particular probability distributions of the noise is not assumed. The main assumptions on the noise terms are that they are independent and have symmetric distributions about zero, however, their distributions can change in each time-step. Regarding regressors, this paper concentrates on the deterministic case, while it is easy to generalize the results to the case where the regressors are random but independent of the noise.

The main contributions of the paper are as follows:
\medskip
\begin{enumerate}
\item The SPS method for building confidence regions for linear regression problems is introduced.
\smallskip
\item The following finite-sample (non-asymptotic) results are proved for SPS under mild statistical assumptions:
\begin{itemize}
\item The probability that the SPS region contains the true parameter is exact for any user-chosen probability.
\item The SPS confidence regions are star convex with the least-squares estimate as a star center.
\end{itemize}
\smallskip
\item The paper also discusses the practical implementation of SPS and introduces an ellipsoidal outer approximation algorithm which can be efficiently computed.
\smallskip
\item Finally, several experiments are presented that illustrate the  SPS method, and compare it with alternatives such as the confidence ellipsoids based on the asymptotic system identification theory.
\end{enumerate}
\medskip
The structure of the paper is as follows. In Section \ref{section-problem} the  linear regression problem is presented together with the assumptions. Section \ref{least-squares} gives a short overview of the LS method and its asymptotic theory. Section \ref{sps-method} introduces the SPS method, while Section \ref{theoretical-results} presents its theoretical properties. In Section \ref{algorithm} the ellipsoidal outer approximation algorithm is described followed by several numerical experiments in Section \ref{simulations}. Finally, Section \ref{conclusions} summarizes and concludes the paper. Preliminary versions of the results in this paper can be found in previous conference papers \cite{Csaji2012a,Csaji2012b,WCC2013}, which also contain additional numerical experiments.

\section{Problem Setting}
\label{section-problem}
This section presents the linear regression problem and introduces our main assumptions and objectives.

\subsection{Data Generation}
Consider the following scalar linear regression system\vspace{1mm}
\begin{equation}
\label{linear-regression}
Y_t \, \triangleq \,
\varphi_t^\mathrm{T} \theta^* + N_t,
\end{equation}
where $Y_t$ is the output, $N_t$ is the noise, $\varphi_t$ is the regressor, and $t$ is the discrete time index. Parameter $\theta^*$ is the true parameter to be estimated. The random variables $Y_t$ and $N_t$ are real-valued, while $\varphi_t$ and $\theta^*$ are  $d$ dimensional real vectors. We consider a finite sample of size $n$ which consists of the regressors $\varphi_1, \dots, \varphi_n$ and the outputs $Y_1, \dots, Y_n$.

For simplicity, we will consider deterministic regressors, $\{\varphi_t\}$, in this paper. Note, however, that our results can be easily generalized to the case of random, but  exogenous, regressors, namely, to the case when the noise sequence $\{N_t\}$ is independent of the regressor sequence $\{\varphi_t\}$. In that case, our assumptions on the regressors (stated below) must be satisfied almost surely and then the analysis can be traced back to the presented theory by fixing a realization of the regressors (i.e., by conditioning on the $\sigma$-algebra generated by the regressors) and applying the presented results realization-wise.

\subsection{Examples}
There are many examples in signal processing and control of systems taking the form of (\ref{linear-regression}) \cite{Ljung1999,Soderstrom1989}. 
The most common example is  the widely used FIR model
\[
 Y_t\,=\,b_1^*U_{t-1}+b^*_2U_{t-2}+\cdots+b^*_dU_{t-d}+N_t,
\]
where $\varphi_t=[U_{t-1},\ldots,U_{t-d}]^\mathrm{T}$ consists of past inputs and $\theta^*=[b_1^*,\ldots,b_d^*]^\mathrm{T}$.

More generally, orthogonal functions (w.r.t.\ the Hardy space $\mathscr{H}_2$) are often used for modeling systems with slowly decaying impulse responses. Their transfer functions can be written as
$$G(z, \theta^*) \, = \, \sum_{k=1}^{d}\theta^*_k\, L_k(z, \alpha),$$
where $z$ is the shift operator and $\{L_k(z, \alpha)\}$ is a function expansion with a (fixed) user-chosen parameter $\alpha$. The regressor in this case is $\varphi_t = [\,L_1(z, \alpha)\,u_t, \ \ldots, \ L_d(z, \alpha)\,u_t\,]^\mathrm{T}$.

Using $L_k(z, \alpha) = z^{-k}$ corresponds to the standard FIR model while, e.g., a Laguerre model is obtained by 
using the Laguerre polynomials \cite{Ljung1999,VanDenHof1995,VanDenHof2005},
$$L_k(z, \alpha)\,=\, \frac{1}{z - \alpha} \left(\frac{1-\alpha z}{z - \alpha} \right)^{k-1}\hspace{-5.5mm}.$$

\subsection{Basic Assumptions}
Our assumptions on the regressors and the noise are:

\medskip
\begin{enumerate}[label=A\arabic*]
\renewcommand{\theenumi}{A\arabic{enumi}}
\item\label{assumption-symmetricindep-noise}{\em (independence, symmetricity): $\{N_t\}$ is a sequence of independent random variables. Each $N_t$ has a symmetric probability distribution about zero.
    }\medskip
\item\label{assumption-outerproduct-invertability}{\em (outer product invertibility): $\det(R_n)\neq 0$, where $$R_n\, \triangleq\, \frac{1}{n}\sum_{t=1}^n\varphi_t \varphi_t^\mathrm{T}.$$}
\end{enumerate}
\medskip

The strongest assumption on the noise is that it forms an independent sequence (see Section \ref{blockSPS} for comments on how this assumption can be relaxed). Apart from independence, the noise assumptions are rather weak, and the noise terms can be nonstationary with unknown distributions, and there are no moment or density requirements either. The other significant assumption is that the noise must be symmetric.
Many standard distributions 
satisfy this property.

\subsection{Objectives}
Our goal is to construct {\em confidence regions} for the parameter $\theta^*$ that have {\em guaranteed} user-chosen confidence probabilities  for finite, and possibly small, number of 
data points. The constructed regions are quasi distribution-free, as the only assumption on the noise distribution is \ref{assumption-symmetricindep-noise}. This is important since in practice the knowledge about the noise distribution is limited. Additionally, the confidence regions should contain the {\em least-squares} point estimate.

We will see that the SPS method proposed in this paper provides finite-sample confidence regions that have an exact user-chosen probability to contain $\theta^*$. Despite the generality of our assumptions, the confidence regions are well-shaped and, in standard cases, they are similar in size to the regions that would be constructed with the full knowledge of the statistical characteristics of the noise.

\section{Least Squares and its Asymptotic Theory}
\label{least-squares}
Before we present the SPS approach, we 
briefly recall the  LS method and its associated asymptotic theory as they are used 
in later sections.

\subsection{Least-Squares Estimate (LSE)}

To find the LSE, we introduce the predictors
$$\hat{Y}_{t}(\theta)\,\triangleq\, \varphi_t^\mathrm{T}\theta.$$
The {prediction errors 
for a given $\theta$ are
$${\varepsilon}_t(\theta)\,\triangleq\,Y_t - \hat{Y}_{t}(\theta)=Y_{t}-\varphi_t^\mathrm{T}\theta,$$
and the LSE is found by minimizing the sum of the squared prediction errors, that is,
$$\hat{\theta}_n\,\triangleq\,
\argmin_{\theta \in \mathbb{R}^d}\sum_{t=1}^n{\varepsilon}^2_t(\theta) =
\argmin_{\theta \in \mathbb{R}^d}\sum_{t=1}^n(Y_t-\varphi_t^\mathrm{T}\theta)^2.$$
The solution can be found by solving the  normal equation,
\begin{equation}
\label{normal-equation}
  \sum_{t=1}^n \varphi_t\, {\varepsilon}_t(\theta) =\sum_{t=1}^n
  \varphi_t(Y_t-\varphi_t^\mathrm{T}\theta)=0,
\end{equation}
which has the analytic solution (assuming \ref{assumption-outerproduct-invertability})
$$
\hat{\theta}_n\,=\, \bigg(\sum_{t=1}^n \varphi_t\varphi_t^\mathrm{T}\bigg)^{\!-1}\bigg( \sum_{t=1}^n \varphi_tY_t\bigg).
$$

\subsection{Asymptotic Confidence Regions}
For zero mean independent and identically distributed (i.i.d.) noise, the LS estimation error is  asymptotically Gaussian under mild conditions. More precisely,  $\sqrt{n}\,({\hat{\theta}_n} - \theta^*)$ converges in distribution to the Gaussian distribution with zero mean and covariance
$\Gamma\, \triangleq\, \sigma^2\, R^{-1}$\!\!\!\!\!\!,\,\,\,\,\,
where $\sigma^2$ is the variance of the noise, and $R$ is the limit of $R_n = \frac{1}{n}\sum_{t=1}^n\varphi_t \varphi_t^\mathrm{T}$  as $n\rightarrow \infty$ assuming this limit exists and is positive definite. As a consequence, $\frac{n}{\sigma^2} ({\hat{\theta}_n} - \theta)^\mathrm{T}\, R\, ({\hat{\theta}_n} - \theta)$ converges in distribution to the $\chi^2$ distribution with $\mbox{dim}(\theta^*) = d$ degrees of freedom \cite{Ljung1999}.

Replacing $R$ with $R_n$ and $\sigma^2$ with the estimate
\begin{equation}
\label{noise-variance-estimate}
\hat{\sigma}^2_n \, \triangleq \, \frac{1}{n-d}\, \sum_{t=1}^{n} {\varepsilon}^2_t(\hat{\theta}_n),
\end{equation}
an approximate confidence region can be built as
\begin{equation}
\label{asymptotic ellipsoid}
\widetilde{\Theta}_n\, \triangleq \, \bigg\{\, \theta \in \mathbb{R}^d\, :\, (\theta - {\hat{\theta}_n})^\mathrm{T}\, R_n\, (\theta - {\hat{\theta}_n}) \, \leq \, \frac{\mu \hat{\sigma}^2_n}{n} \,\bigg\},
\end{equation}
where the probability that $\theta^*$ is in the confidence region $\widetilde{\Theta}_n$ is  approximately $F_{\chi^2(d)}(\mu)$, where $F_{\chi^2(d)}$ is the cumulative distribution function of the $\chi^2$ distribution with $d$ degrees of freedom. However, this confidence region based on the asymptotic system identification theory does not come with rigorous theoretical guarantees for the finite sample case, and therefore should only be used as a heuristic.

\section{The Sign-Perturbed Sums (SPS) Method}
\label{sps-method}
In this section, we first motivate, and then formally introduce, the {\em Sign-Perturbed Sums} (SPS) method for constructing confidence regions with guaranteed finite sample properties.

\subsection{Intuitive Idea}
The LS estimate is obtained as the solution to the normal equation (\ref{normal-equation}). This equation can be re-written as
$$
\sum_{t=1}^n \varphi_t\varphi_t^\mathrm{T} \tilde{\theta}+ \sum_{t=1}^n \varphi_t N_t = 0,
$$
where $\tilde{\theta}\, \triangleq\, \theta^* - \theta$. It is clear that the uncertainty in the LSE comes from the noise $\{N_t\}$, and, if each $N_t$ were zero, then $\hat{\theta}_{n} = \theta^*$. In order to construct a confidence region, we should somehow evaluate the uncertainty of the estimate. One way of doing this would be to assume a particular probability distribution of the noise  and propagate this distribution through the above formula to get a distribution of the estimation error. Then, the distribution of the estimation error can be used to construct the confidence region. However, we want to avoid such an approach as it needs strong prior assumptions on the noise, which makes it unattractive for practical purposes.

We follow another approach and try to exploit the information in the data as much as possible while assuming minimal prior statistical knowledge about the noise. Our core assumption is the {\em symmetry} of the noise. 
We introduce $m-1$ {\em sign-perturbed sums}
\begin{eqnarray*}
H_i(\theta)& = & \sum_{t=1}^n  \varphi_t\alpha_{i,t}(Y_t-\varphi_t^\mathrm{T}\theta)  \\ &=& \sum_{t=1}^n \alpha_{i,t}\varphi_t\varphi_t^\mathrm{T} \tilde{\theta}+ \sum_{t=1}^n\alpha_{i,t} \varphi_t N_t, \ \ \
\end{eqnarray*}
$ i=1,\ldots, m-1$, 
where $\{\alpha_{i,t}\}$ are random signs, i.e.\ i.i.d.\ random variables that take on the values $\pm 1$  with probability 1/2 each. That is, we perturb the sign of the prediction errors in the normal equation. 
For a given $\theta$, we can also calculate the value for the case when no sign-perturbations are used, which we call the {\em reference sum},
$$
H_0(\theta) = \sum_{t=1}^n \varphi_t(Y_t-\varphi_t^\mathrm{T}\theta) = \sum_{t=1}^n \varphi_t\varphi_t^\mathrm{T} \tilde{\theta}+ \sum_{t=1}^n \varphi_t N_t.
$$
A comparison of the $H_0(\theta)$ and $H_i(\theta)$ functions can be done by using a norm $\|\cdot\|$. In the sequel, if not otherwise stated, $\|\cdot\|$ will refer to the 2-norm, i.e. $\|x\|^2= x^\mathrm{T}x$. 

For $\theta = \theta^*$, these sums can be simplified to
\begin{eqnarray*}
H_0(\theta^*)  &\!\!\!\!=\!\!\!\!& \sum_{t=1}^n \varphi_t N_t,\\
H_i(\theta^*)  &\!\!\!\!=\!\!\!\!& \sum_{t=1}^n \alpha_{i,t} \varphi_t N_t=\sum_{t=1}^n \pm \varphi_t N_t,
\end{eqnarray*}
where in the last equation we have written  $\pm$ instead of $\alpha_{i,t}$ for intuitive understanding. $H_0(\theta^*)$ and $H_i(\theta^*)$ have the {\em same distribution} since $\{N_t\}$ are independent and symmetric. Therefore, there is no reason why a particular $\|H_{j}(\theta^*)\|^2$ should be bigger or smaller than another $\|H_i(\theta^*)\|^2$ and the probability that a particular $\|H_{j}(\theta^*)\|^2$ is the $k\hspace{0.2mm}$th largest one in the ordering of $\{\|H_i(\theta^*)\|^2\}_{i=0}^{m-1}$ will be the same for all $j$, including $j=0$ (the case of the reference sum, i.e., where there are no sign-perturbations). As $j$ can take on $m$ different values, this probability is exactly $1/m$.

However, for ``{\em large enough}'' $\|\tilde{\theta}\|$, we will have that
$$
\bigg\|\sum_{t=1}^n \varphi_t\varphi_t^\mathrm{T} \tilde{\theta} + \sum_{t=1}^n \varphi_t N_t\,\bigg\|^2 > \bigg\|\sum_{t=1}^n \pm \varphi_t\varphi_t^\mathrm{T} \tilde{\theta} + \sum_{t=1}^n \pm \varphi_t N_t\,\bigg\|^2,
$$
with ``{\em high probability}''. In fact,  $\sum_{t=1}^n \varphi_t\varphi_t^\mathrm{T} \tilde{\theta}$ on the left-hand side  increases faster than $\sum_{t=1}^n \pm \varphi_t\varphi_t^\mathrm{T} \tilde{\theta}$ on the right-hand side. Hence, for $\|\tilde{\theta}\|$ large enough, $\|H_0(\theta)\|^2$ dominates in the ordering of $\{\|H_i(\theta)\|^2\}$.

From these intuitions, the general idea is to construct the confidence region based on the {\em rankings} of the functions $\{\|H_i(\theta)\|^2\}$ and leave out those $\theta$ parameters for which $\|H_0(\theta)\|^2$ ``{\em dominates}'' the other functions. 

In the formal construction of the SPS method, functions $\{H_i(\theta)\}$ will be modified with a term, $R_n^{-1/2}$, that helps to shape the region and an $1/n$ factor to increase the numerical stability, that is, we will use $S_i(\theta) = R_n^{-1/2}\frac{1}{n}H_i(\theta)$ instead of $H_i(\theta)$, cf.\ the pseudocode in Table \ref{indtab}. However, this does not affect the core idea of the construction. 
Next, we provide the formal construction of SPS, followed by  results stating some finite-sample properties of the obtained confidence sets.

\subsection{Confidence Region Construction}

The SPS method is in two parts. The first part, which we call initialization, sets the main global parameters of SPS and generates the random objects needed for the construction. In the initialization, the user provides the desired confidence probability $p$. The second part evaluates an indicator function, which can be called for a particular parameter value $\theta$ to decide whether it is included in the confidence region.

The pseudocode for the initialization is  given in Table \ref{inittab}.
{\renewcommand{\arraystretch}{1.3}
\begin{table}[h]
\normalsize
\vspace*{1mm}
\begin{center}
\begin{tabular}{|rlll|}
\hline
\multicolumn{4}{|c|}{\scshape Pseudocode: SPS-Initialization} \\
\hline \hline 1. & \multicolumn{3}{l|}{Given a (rational) confidence probability $p \in (0,1)$,} \\
 & \multicolumn{3}{l|}{set integers $m > q > 0$ such that $p = 1 - q/m$;}\\
2. & \multicolumn{3}{l|}{Calculate the outer product} \\
& \multicolumn{3}{c|}{$R_n\, \triangleq\, \frac{1}{n}\sum\limits_{t=1}^n\varphi_t \varphi_t^\mathrm{T}$,}\\
& \multicolumn{3}{l|}{ and find a factor $R_n^{1/2}$ such that} \\
& \multicolumn{3}{c|}{$R_n^{1/2}R_n^{1/2\mathrm{T}}=R_n$;}\\
3. & \multicolumn{3}{l|}{Generate $n\,(m-1)$ i.i.d.\ random signs $\{\alpha_{i,t}\}$ with} \\
& \multicolumn{3}{c|}{$\mathbb{P}(\alpha_{i,t} = 1)\, = \,\mathbb{P}(\alpha_{i,t} = -1) \,=\, \frac{1}{2}$,}\\
& \multicolumn{3}{l|}{for $i \in \{1, \dots, m-1\}$ and $t \in \{1, \dots, n\}$;}\\
4. & \multicolumn{3}{l|}{Generate a random permutation $\pi$ of the set}\\
& \multicolumn{3}{l|}{$\{0, \dots, m-1\}$, where each of the $m!$ possible}\\
& \multicolumn{3}{l|}{permutations has the same probability $1/(m!)$}\\
& \multicolumn{3}{l|}{to be selected.}\\
\hline
\end{tabular}
\end{center}
\caption{}\label{inittab}
\end{table}}
The permutation $\pi$ in point 4 is only used in the indicator function to break ties, and decide which function $||S_i(\theta)||^2$ or $||S_j(\theta)||^2$ is the ``{\em larger}'' if $||S_i(\theta)||^2$ and $||S_j(\theta)||^2$ take on the same value. More precisely, given $m$ real numbers $\{Z_i\}$, $i = 0,\ldots,m-1$, we define a strict total order $\succ_{\pi}$ by
$$Z_k \succ_{\pi} Z_j \hspace{10mm}\hbox{if and only if}$$
$$\left(\,Z_k > Z_j\,\right) \hspace{2mm}\hbox{or}\hspace{2mm} \left(\,Z_k = Z_j \hspace{2mm}\hbox{and}\hspace{2mm} \pi(k) > \pi(j)\,\right).$$

Note that $\pi$ is a bijection (one-to-one correspondence) from $\{0, \dots, m-1\}$ to itself, thus, for $k \neq j$, $\pi(k)$ and $\pi(j)$ are two different integers in $\{0, \dots, m-1\}$.

After SPS is initialized, the indicator function  given in Table \ref{indtab} can be called to decide whether a particular parameter value $\theta$ is included in the confidence region.\\

{\renewcommand{\arraystretch}{1.3}
\begin{table}[h]
\normalsize
\begin{center}
\begin{tabular}{|rlll|}
\hline
\multicolumn{4}{|c|}{\scshape Pseudocode: SPS-Indicator\,(\,$\theta$\,)}\\
\hline \hline 1. & \multicolumn{3}{l|}{For the given $\theta$, compute the prediction errors }\\
& \multicolumn{3}{l|} {for $t \in \{1, \dots, n\}$} \\
 & \multicolumn{3}{c|}{${\varepsilon}_t(\theta)\,\triangleq\, Y_t - \varphi_t^\mathrm{T}\theta$;}\\
2. & \multicolumn{3}{l|}{Evaluate} \\
& \multicolumn{3}{l|}{\hspace{15mm}$S_0(\theta) \triangleq R_n^{-\frac{1}{2}} \frac{1}{n}\sum\limits_{t=1}^{n}{\, \varphi_t {\varepsilon}_t(\theta)}$,}\\
& \multicolumn{3}{l|}{\hspace{15mm}$ S_i(\theta) \triangleq R_n^{-\frac{1}{2}} \frac{1}{n}\sum\limits_{t=1}^{n}{\, \alpha_{i,t} \, \varphi_{t}{\varepsilon}_t(\theta)}$,}\\
& \multicolumn{3}{l|}{for $i \in \{1, \dots, m-1 \}$;}\\
3. & \multicolumn{3}{l|}{Order scalars $\{\|S_i(\theta)\|^2\}$ according to $\succ_{\pi}$;}\\
4. & \multicolumn{3}{l|}{Compute the rank $\mathcal{R}(\theta)$ of $\|S_0(\theta)\|^2$ in the ordering,} \\
& \multicolumn{3}{l|} {where $\mathcal{R}(\theta) = 1$ if $\|S_0(\theta)\|^2$ is the smallest in the} \\
& \multicolumn{3}{l|} {ordering, $\mathcal{R}(\theta) = 2$ if $\|S_0(\theta)\|^2$ is the second} \\
& \multicolumn{3}{l|} {smallest, and so on.}\\
6. & \multicolumn{3}{l|}{Return $1$ if $\mathcal{R}(\theta) \leq m-q$, otherwise return $0$.}\\
\hline
\end{tabular}
\end{center}
\caption{}\label{indtab}
\vspace{-4mm}
\end{table}}

Using this construction, we can define the $p$-level {\em SPS confidence region} as follows
\begin{equation*}
\widehat{\Theta}_n \, \triangleq \, \left\{\, \theta \in \mathbb{R}^d\, :\, \text{SPS-INDICATOR}(\,\theta\,) = 1\, \right\}.
\end{equation*}

Observe that the LS estimate, $\hat{\theta}_{n}$, has by definition the property that $S_0(\hat{\theta}_{n})=0$. Therefore, the LSE is included in the SPS confidence region, assuming that it is non-empty.\footnote{There is a positive, but negligible, probability that the SPS confidence region is empty. This happens, if for at least $m-q$ indices $i$, the $\{\alpha_{i,t}\}$ sequences are sequences of all  $+1$s or all $-1$s, and  $m-q$ or more of the corresponding $\{S_i\}$ functions are ranked smaller than $S_0$ by $\pi$.}

The SPS method in the form developed in this section lends itself nicely to problems where the indicator function should only be evaluated for finitely many values of $\theta$. This happens for example in certain hypothesis testing or change detection problems. Later, we will discuss ways to make SPS suitable for problems where one wishes to represent the whole SPS confidence regions compactly.

\section{Theoretical Results}
\label{theoretical-results}

\subsection{Exact Confidence}\label{exact_confidence section}
The most important property of the SPS method is that the regions it generates have {\em exact} confidence probabilities for any {\em finite} sample. 
The following theorem holds.
\medskip
\begin{theorem}\label{theorem-exact-prob}
 {\em Assuming \ref{assumption-symmetricindep-noise} and \ref{assumption-outerproduct-invertability}, the confidence probability of the constructed confidence region is exactly $p$, that is,}
\begin{equation*}
\mathbb{P}\big(\theta^* \in \widehat{\Theta}_n\big)\, =\, 1 - \frac{q}{m} \, = \, p.
\end{equation*}
\end{theorem}
\medskip
 A formal proof of Theorem \ref{theorem-exact-prob} can be found in Appendix \ref{appendix-exactness-proof}. Interestingly, the proof does not depend on the applied norm, and the result keeps its validity regardless of the norm used in step 3 in the SPS indicator function when constructing $\widehat{\Theta}_n$.
Since the confidence probability is exact, no conservatism is introduced. Moreover, the statistical assumptions imposed on the noise are mild, e.g., knowledge of the particular noise distribution is not assumed, the noise can change in each time step, and there are no moment or density assumptions.

The simulation examples in Section \ref{simulations} also demonstrate that, when the noise is stationary, the SPS confidence regions compare in size with the approximate confidence regions obtained by applying the asymptotic system identification theory, while, unlike asymptotic regions, the SPS regions  maintain their guaranteed validity even for nonstationary noise patterns.

\subsection{Star Convexity}
Earlier we observed that the LSE is in the SPS confidence region. The next theorem makes our claim that the SPS regions are built around the LS estimate more precise.

Recall that set $\mathcal{X} \subseteq \mathbb{R}^d$ is called {\em star convex} if there is a {\em star center} $c \in \mathbb{R}^d$, such that
$$\forall x \in \mathcal{X}, \forall\,\beta \in [0,1]: \beta\,x + (1-\beta)\, c \in \mathcal{X}.$$
All convex sets are
star convex, but the converse is not true.

It is easy to construct examples that show that, in general, the SPS confidence regions are not convex. For example, if $q=1$, the SPS region is the union of ellipsoids, and it is typically non-convex. On the other hand, as the next theorem demonstrates, the SPS confidence regions are star convex.
\medskip
\begin{theorem}\label{theorem-starset}
 {\em Assuming \ref{assumption-symmetricindep-noise} and \ref{assumption-outerproduct-invertability}, the SPS confidence regions are star convex with the LS estimate as a star center.}
\end{theorem}
\medskip
The proof of Theorem \ref{theorem-starset} is given in Appendix \ref{appendix-starset-proof}.

This result not only shows that the SPS regions are centered around the LS estimate, but it also provides a basis for finding the boundary of the SPS region. In fact, one can search rays from the LS estimate outwards for the first point which is not in the SPS region, and by the star convexity property this will be a boundary point of the SPS region. 

\subsection{A More General Algorithm: Block SPS}\label{blockSPS}
The fundamental assumption regarding the noise terms is that they are symmetric about zero and independent. Theoretically, it is easy to relax the independence assumption and allow dependent noises as long as their signs are independent. Moreover, robustness against the independence assumption can be boosted by using a modified SPS method where the random signs $\{\alpha_{i,t}\}$ are kept at the same value $+1$ or $-1$ for blocks of $T$ consecutive time instants before the sign is again randomly drawn. The only difference is in point 3 of the SPS-Initialization algorithm, which now becomes (assuming, for simplicity, that $n/T$ is an integer)
{\renewcommand{\arraystretch}{1.3}
\begin{table}[h]
\normalsize
\begin{center}
\begin{tabular}{|rlll|}\hline
3'. & \multicolumn{3}{l|}{Generate $\frac{n}{T}\,(m-1)$ i.i.d.\ random signs $\{\bar{\alpha}_{i,k}\}$ with} \\
& \multicolumn{3}{c|}{$\mathbb{P}(\bar{\alpha}_{i,k} = 1)\, = \,\mathbb{P}(\bar{\alpha}_{i,k} = -1) \,=\, \frac{1}{2}$,}\\
& \multicolumn{3}{l|}{for $i \in \{1, \dots, m-1\}$, $k \in \{1, \dots, n/T\}$, and let}\\
& \multicolumn{3}{c|}{$\alpha_{i,(k-1)T+j}=\bar{\alpha}_{i,kT}$}\\
& \multicolumn{3}{l|}{for $i \in \{1, \dots, m-1\}$, $k \in \{1, \dots, n/T\}$, and}\\
& \multicolumn{3}{l|}{$j\in\{1,\ldots T\}$.}\\ \hline
\end{tabular}
\vspace{-2mm}
\end{center}
\end{table}}

If the noise is dependent and $T$ is larger than the dominant time constant of the noise dynamics, then the blocks of $T$ consecutive noise terms act approximately as independent noise terms so that the result in Theorem \ref{theorem-exact-prob} holds approximately.

When instead this modified Block SPS method is applied to systems where the noise is actually independent, the exact confidence result in Theorem \ref{theorem-exact-prob} remains  valid as can be seen from an inspection of the proof. Moreover, for independent noise, the regions constructed by SPS and Block SPS methods are not very different and Block SPS is only marginally worse. See Section \ref{blockSPSsim} for a simulation example.

\subsection{Comparison to Bootstrap}
A common feature in SPS and bootstrap approaches is that randomization is an essential ingredient.

In case of linear regression problems \cite{Efron1993,davidson2006,Godfrey2009}, we are interested in the distribution of the noise, however, we do not have a direct access to the noise samples. In order to overcome this difficulty, bootstrap typically uses the prediction errors and works with the sample of residuals, $\{\varepsilon_t(\theta)\}$, as an estimate of the noise, instead of the sample of the (unobserved) noise terms, $\{N_t\}$. One can use residuals for different parameters to test various hypotheses \cite{Godfrey2009} or, as is common,  one can work with the prediction errors of a nominal estimate \cite{Efron1993,davidson2006}, such as the LS residuals, $\{\varepsilon_t(\hat{\theta}_n)\}$.

The latter case corresponds to work with the new data set(s) $\{\varphi_{\kappa}, Y'_{\kappa}\}$  where $Y'_{\kappa}$ is generated by
$Y'_{\kappa} = \varphi_{\kappa}^\mathrm{T}\hat{\theta}_n + \tilde{\varepsilon}_{\kappa}$,
with $\tilde{\varepsilon}_{\kappa}$
randomly selected (uniformly, with replacement) from $\{\varepsilon_t(\hat{\theta}_n)\}$.
Various statistics can then be calculated from the resampled data set(s). An alternative way of generating data set(s) is pairs bootstrap \cite{davidson2006}, where data sets are generated by random selection (with replacement) of regressor-output pairs, $(\varphi_{\tau}, Y_{\tau})$, and the bootstrap samples are built from these pairs.

The domain of applicability of bootstrap is larger than SPS, since there is no assumption that the noise is symmetric and there are also bootstrap methods that can handle correlated noise. Moreover, bootstrap can be applied to non-linear models. On the other hand, while there are theoretical asymptotic results for bootstrap methods, there are few finite sample results and hence the results based on bootstrap are in most cases only approximate in the finite sample case. For example, if the approach for estimating the covariance matrix of the LS estimate, found in Ch.9 of Efron and Tibshirani's classical book \cite{Efron1993}, is combined with an assumption that the estimate has a Gaussian distribution, then the confidence ellipsoids of asymptotic system identification theory are obtained and they are approximate and not exact in a finite sample setting.

\section{Ellipsoidal Approximation Algorithm}
\label{algorithm}

Given a particular value of $\theta$, it is easy to check whether $\theta$ is in the confidence region. All we have to do  is to 
calculate the $\{\|S_i(\theta)\|^2\}$ functions for that $\theta$ and compare them. 
Hence the SPS confidence regions can be constructed by checking each parameter value on a grid. However, this approach is computationally demanding and suffers from the ``{\em curse of dimensionality}''. Here, we present an approximation algorithm for SPS that can be efficiently computed (i.e., in polynomial time) and offers a compact representation in the form of ellipsoidal over-bounds. An alternative approach based on interval analysis has also been proposed \cite{KiefferWalter2013,KiefferWalter2013a}.

\subsection{Ellipsoidal Outer Approximation}
Expanding $\|S_0(\theta)\|^2$, we find that it can be written as
\begin{align*}
  \|S_0(\theta)\|^2\! &=\!\bigg[\frac{1}{n}\!\sum_{t=1}^n\varphi_t(Y_t-\varphi_t^\mathrm{T}\theta)\bigg]^{\!\mathrm{T}}\!\!R_n^{-1}
  \bigg[\frac{1}{n}\!\sum_{t=1}^n\varphi_t(Y_t-\varphi_t^\mathrm{T}\theta)\bigg] \\ & =\!\bigg[\frac{1}{n}\!\sum_{t=1}^n\varphi_t\varphi_t^\mathrm{T}(\theta-\hat{\theta}_n)\bigg]^\mathrm{T}\!\!\!R_n^{-1} \bigg[\frac{1}{n}\!\sum_{t=1}^n\varphi_t\varphi_t^\mathrm{T}(\theta-\hat{\theta}_n)\bigg] \\ & = (\theta-\hat{\theta}_n)^{\mathrm{T}}R_n(\theta-\hat{\theta}_n),
  \end{align*}

For the purpose of finding an ellipsoidal over-bound we can ignore the random ordering used when $||S_0(\theta)||^2$ and $||S_i(\theta)||^2$ are equal, and consider the set   given by those values of $\theta$  at which $q$ of the $||S_i(\theta)||^2$ are larger {\em or equal} to $||S_0(\theta)||^2$, i.e.,
\begin{equation*}
\widehat{\Theta}_n \, \subseteq \, \left\{\, \theta \in \mathbb{R}^d\, :\, (\theta-\hat{\theta}_n)^\mathrm{T}R_n(\theta-\hat{\theta}_n)\leq r(\theta) \, \right\},
\end{equation*}
where $r(\theta)$ is the $q$\hspace{0.3mm}th largest value of functions $\{\|S_i(\theta)\|^2\}$, $i=1,\ldots,m-1$.
The idea is now to seek an  over-bound by replacing $r(\theta)$ with a parameter independent $r$. This outer approximation will hence have the {\em same shape} and {\em orientation} as the asymptotic  confidence ellipsoid (\ref{asymptotic ellipsoid}), but it will have a different volume. The outer approximation is a guaranteed confidence region for finitely many data points. Moreover, it will have a  compact representation, since it is characterized in terms of $\hat{\theta}_n, R_n$ and $r$.

\subsection{Convex Programming Formulation}
Comparing $\|S_0(\theta)\|^2$ with one single $\|S_i(\theta)\|^2$ function, we have
\begin{eqnarray}
\lefteqn{
\{\,
\theta: \|S_0(\theta)\|^2 \leq \|S_i(\theta)\|^2
\,\}
} \nonumber \\
& & \subseteq
\{\,
\theta: \|S_0(\theta)\|^2 \leq \max_{\theta: \|S_0(\theta)\|^2 \leq \|S_i(\theta)\|^2}\|S_i(\theta)\|^2
\,\}. \nonumber
\end{eqnarray}
Relation $\|S_0(\theta)\|^2 \leq \|S_i(\theta)\|^2$ can be rewritten as
$$
(\theta-\hat{\theta}_{n})^\mathrm{T}R_n(\theta-\hat{\theta}_{n}) \leq
$$
$$
\bigg[\frac{1}{n}\sum_{t=1}^n\alpha_{i,t}\varphi_t(Y_t-\varphi^\mathrm{T}_t\theta)\bigg]^{\!\mathrm{T}}\!\!R_n^{-1}\!\bigg[\frac{1}{n}\sum_{t=1}^n\alpha_{i,t}\varphi_t(Y_t-\varphi^\mathrm{T}_t\theta) \bigg]
$$
\begin{equation*}
= \,\theta^\mathrm{T}Q_{i}R_n^{-1}Q_{i}\theta-2\,\theta^\mathrm{T}Q_{i}R^{-1}_n{\psi}_{i}+\psi^{\mathrm{T}}_{i}R^{-1}_n{\psi}_{i},
\label{constr1}
\end{equation*}
where matrix $Q_i$ and vector $\psi_i$ are defined as
\begin{eqnarray*}
Q_{i}&\triangleq&\frac{1}{n}\sum_{t=1}^n\alpha_{i,t}\varphi_t\varphi^\mathrm{T}_t, \label{Qieq}\\
\psi_{i} &\triangleq&\frac{1}{n}\sum_{t=1}^n\alpha_{i,t} \varphi_t Y_t. \label{psieq}
\end{eqnarray*}
Noting that 
$$\max_{\theta: \|S_0(\theta)\|^2 \leq \|S_i(\theta)\|^2}\|S_i(\theta)\|^2=\max_{\theta: \|S_0(\theta)\|^2 \leq \|S_i(\theta)\|^2}\|S_0(\theta)\|^2$$ and using the notation $z\,\triangleq\,R_n^{\frac{1}{2}\mathrm{T}}(\theta-\hat{\theta}_n)$, the quantity $\max_{\theta: \|S_0(\theta)\|^2 \leq \|S_i(\theta)\|^2}\|S_i(\theta)\|^2$ can be obtained as the value of the following optimization problem
\begin{eqnarray}
\label{opt1}
&  \mathrm{maximize} & \|z\|^2 \nonumber \\
& \mbox{subject to } & z^\mathrm{T}A_{i}z+2 z^\mathrm{T}b_{i}+c_{i}\leq 0, \nonumber \\
\vspace{-1mm}
\end{eqnarray}
where $A_{i}$, $b_{i}$ and $c_{i}$ are defined as
\begin{eqnarray*}
A_{i}&\triangleq&I-R_n^{-\frac{1}{2}}Q_{i}R_n^{-1}Q_{i}R_n^{{-\frac{1}{2}\mathrm{T}}},\\
b_{i}&\triangleq&R^{-\frac{1}{2}}_n Q_{i} R_n^{-1}(\psi_i-Q_i\hat{\theta}_n),\\
c_{i}&\triangleq&-\psi_i^\mathrm{T} R_n^{-1}\psi_i+2\hat{\theta}^\mathrm{T}_n Q_i R_n^{-1}\psi_i- \hat{\theta}^\mathrm{T}_n Q_i R_n^{-1} Q_i \hat{\theta}_n.
\end{eqnarray*}
This program is not convex in general. However, it can be shown
\cite[Appendix B]{Boyd2009} that {\em strong duality} holds, so that the value of the above optimization program is equal to the value of its dual which can be formulated as
\begin{eqnarray}
\label{opt1-dual}
&  \mathrm{minimize} & \gamma \nonumber \\
& \mbox{subject to } &
\lambda\geq 0 \nonumber \\
& &
\left[\begin{array}{cc} -I+\lambda A_{i} & \lambda b_{i} \\ \lambda b_{i}^\mathrm{T} & \lambda c_{i} + \gamma\end{array}\right]\succeq 0, \nonumber \\
\vspace{-1mm}
\end{eqnarray}
where ``$\succeq 0$'' denotes that a matrix is positive semidefinite. This program is convex, and can be easily solved using, e.g., Yalmip \cite{Lovberg2004} and a solver such as SDPT3.

Letting $\gamma_i^*$ be the value of program \eqref{opt1-dual}, we now have
$$
\{
\theta: \|S_0(\theta)\|^2 \leq \|S_i(\theta)\|^2
\}
\subseteq
\{
\theta: \|S_0(\theta)\|^2 \leq \gamma_i^*
\}.
$$
Thus,
\begin{equation*}
\widehat{\Theta}_n \subseteq
\doublehat{\Theta}_n \, \triangleq \, \left\{\, \theta \in \mathbb{R}^d\, :\, (\theta-\hat{\theta}_n)^\mathrm{T}R_n(\theta-\hat{\theta}_n)\leq r \, \right\},
\end{equation*}
where $r = q\hspace{0.2mm}$th largest value of $\gamma_i^*$, $i=1,\ldots,m-1$.

$\doublehat{\Theta}_n$ is the sought outer approximation. It is clear that
\begin{equation*}
\mathbb{P}\big(\,\theta^* \in \doublehat{\Theta}_n\,\big)\, \geq\, 1 - \frac{q}{m} \, = \, p,
\end{equation*}
for any finite $n$.

The pseudocode for computing $\doublehat{\Theta}_n$ is given in Table \ref{approxtab}.

\medskip
{\renewcommand{\arraystretch}{1.3}
\begin{table}[h]
\normalsize
\vspace*{1mm}
\begin{center}
\begin{tabular}{|rlll|}
\hline
\multicolumn{4}{|c|}{\scshape Pseudocode: SPS-Outer-Approximation}\\
\hline \hline 1. & \multicolumn{3}{l|}{Compute the least-squares estimate,} \\
 & \multicolumn{3}{c|}{$\hat{\theta}_n=R_n^{-1}\bigg[\frac{1}{n}\sum\limits_{t=1}^n\varphi_t Y_t\bigg]$;}\\
2. & \multicolumn{3}{l|}{For $i \in \{1, \dots, m-1\}$, solve the optimization} \\
& \multicolumn{3}{l|}{problem (\ref{opt1}), and let $\gamma_i^*$ be the optimal value;}\\
3. & \multicolumn{3}{l|}{Let $r$ be the $q$\hspace{0.3mm}th largest $\gamma_i^*$ value;}\\
4. & \multicolumn{3}{l|}{The outer approximation of the SPS confidence}  \\ & \multicolumn{3}{l|}{region is given by the ellipsoid} \\
& \multicolumn{3}{l|}{$\doublehat{\Theta}_n =\big\{\, \theta \in \mathbb{R}^d\, :\, (\theta-\hat{\theta}_n)^\mathrm{T}R_n(\theta-\hat{\theta}_n)\leq r \, \big\}$.}\\
\hline
\end{tabular}
\end{center}
\caption{}\label{approxtab}
\vspace*{-5mm}
\end{table}}

\section{Simulation examples}
\label{simulations}

In this section we illustrate SPS with numerical examples. The confidence regions constructed by SPS are compared with those obtained using asymptotic system identification theory, and, when the noise is i.i.d. Gaussian, with the ellipsoids based on the $F$-distribution.
The effect of using norms other than the 2-norm as well as the presence of unmodelled dynamics are also studied. The block SPS algorithm is illustrated on an example where the assumptions on the noise are not satisfied.

\subsection{Second Order FIR System}\label{secfir}

We consider a second order data generating FIR system
\[
  Y_t\,=\,b^*_1U_{t-1}+b^*_2U_{t-2}+N_t,
\]
where $b^*_1=0.7$ and $b^*_2=0.3$ are the true system parameters and $\{N_t\}$ is a sequence of i.i.d.\ Laplacian random variables with zero mean and variance 0.1. The input signal is given by
\[
 U_t\,=\,0.75\,U_{t-1}+V_t,
\]
where $\{V_t\}$ is a sequence of i.i.d.\
Gaussian random variables with zero mean and variance $1$.

The model class is the class of second order FIR models, and hence the predictor is given by
\[
 \hat{Y}_t(\theta)\,=\,b_1U_{t-1}+b_2U_{t-2} = \varphi_t^\mathrm{T} \theta,
\]
where $\theta=[\,b_1, b_2\,]^\mathrm{T}$ is the model parameter, and $\varphi_t = [\, U_{t-1}, U_{t-2} \,]^\mathrm{T}$.

Based on $n=25$ data points $(\varphi_t,Y_t) = ([\, U_{t-1}, \ U_{t-2} \,]^\mathrm{T},Y_t)$, $t=1,\ldots,25$, we want to find a 95\% confidence region for $\theta^*=[b^*_1,\ b^*_2]^\mathrm{T}$.
Following the SPS procedure we first compute the matrix
\[
 R_{25}=\frac{1}{25}\sum_{t=1}^{25}\left[\begin{array}{c} U_{t-1} \\ U_{t-2} \end{array}\right]\left[ U_{t-1}, \ \ U_{t-2}\right],
\]
and find a factor $R_{25}^{\frac{1}{2}}$ such that $R_{25}^{\frac{1}{2}}R_{25}^{\frac{1}{2}\mathrm{T}}=R_{25}$.

Then we compute the reference sum
\[
  S_0(\theta)=R_{25}^{-\frac{1}{2}}\frac{1}{25}\sum_{t=1}^{25}\left[\begin{array}{c} U_{t-1} \\ U_{t-2} \end{array}\right](Y_t-b_1U_{t-1}-b_2U_{t-2}),
\]
and the 99 sign perturbed sums, $i=1,\ldots,99$,
\[
 S_i(\theta)=R_{25}^{-\frac{1}{2}}\frac{1}{25}\sum_{t=1}^{25} \alpha_{i,t} \left[\begin{array}{c} U_{t-1} \\ U_{t-2} \end{array}\right] (Y_t-b_1U_{t-1}-b_2U_{t-2}),
\]
where $\alpha_{i,t}$ are i.i.d.\ random signs. Moreover, we generate a random permutation $\pi$ to break possible ties.

\begin{figure}[t]
\vspace*{-1mm}
\setlength{\epsfysize}{6.0cm}\setlength{\epsfxsize}{8.0cm}
\begin{center}
\leavevmode \epsfbox{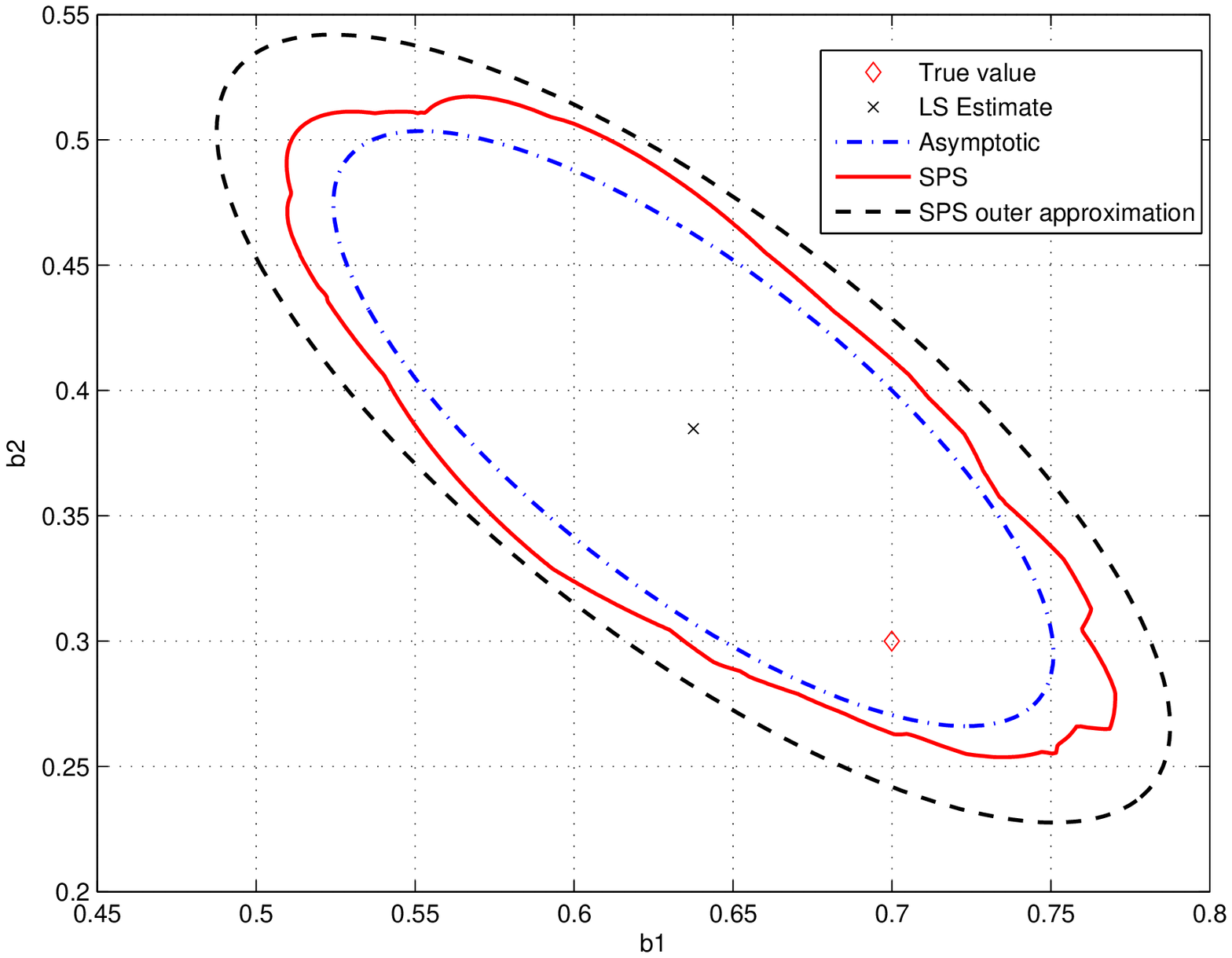}
\vspace{-5mm}
\end{center}\caption{$95$\% confidence regions, $n=25$, $m=100$.}\label{N25}
\medskip
\setlength{\epsfysize}{6.0cm} \setlength{\epsfxsize}{8.0cm}
\begin{center}
\leavevmode \epsfbox{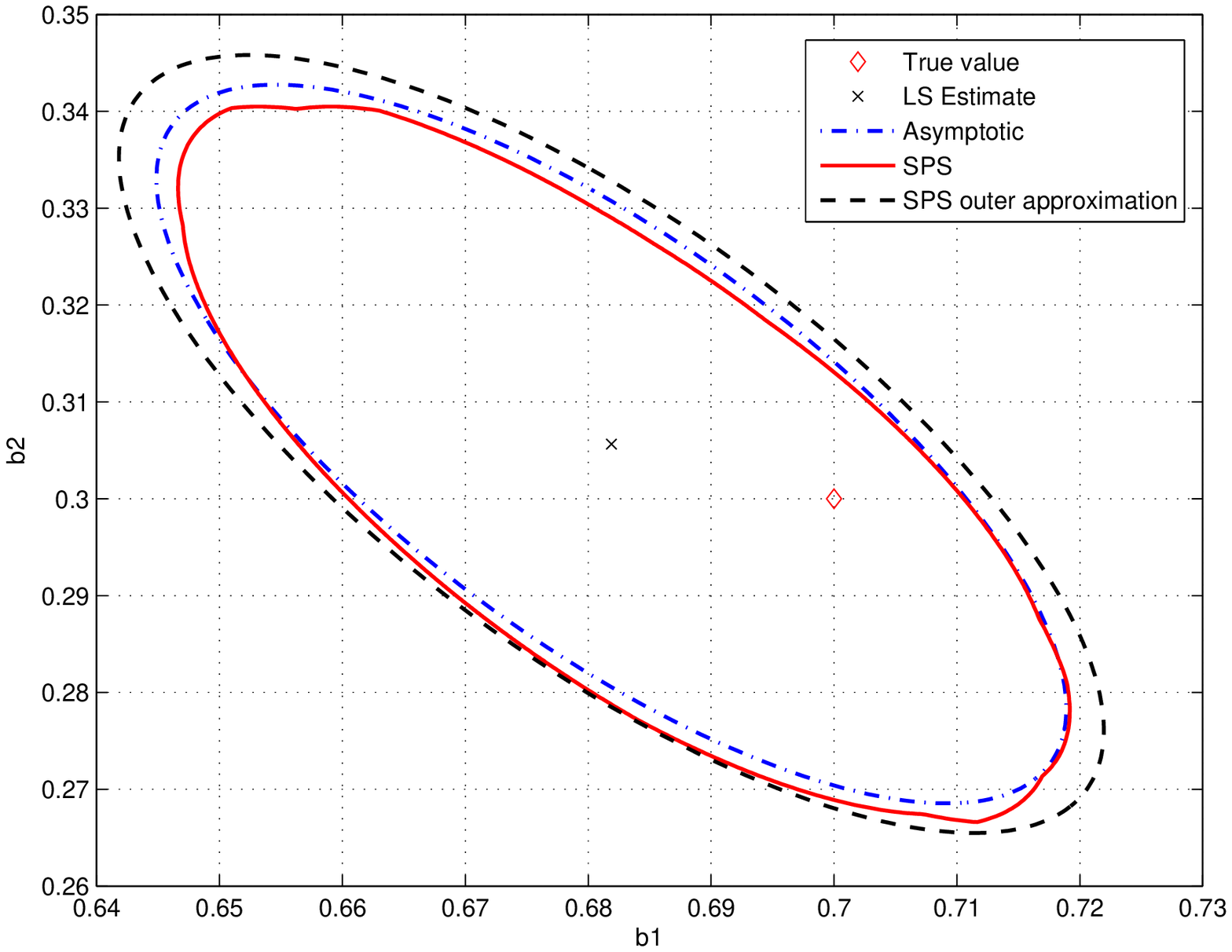}
\vspace{-5mm}
\end{center}\caption{$95$\% confidence regions, $n=400$, $m=100$.}\label{N400}
\medskip
\setlength{\epsfysize}{6.0cm} \setlength{\epsfxsize}{8.0cm}
\begin{center}
\leavevmode \epsfbox{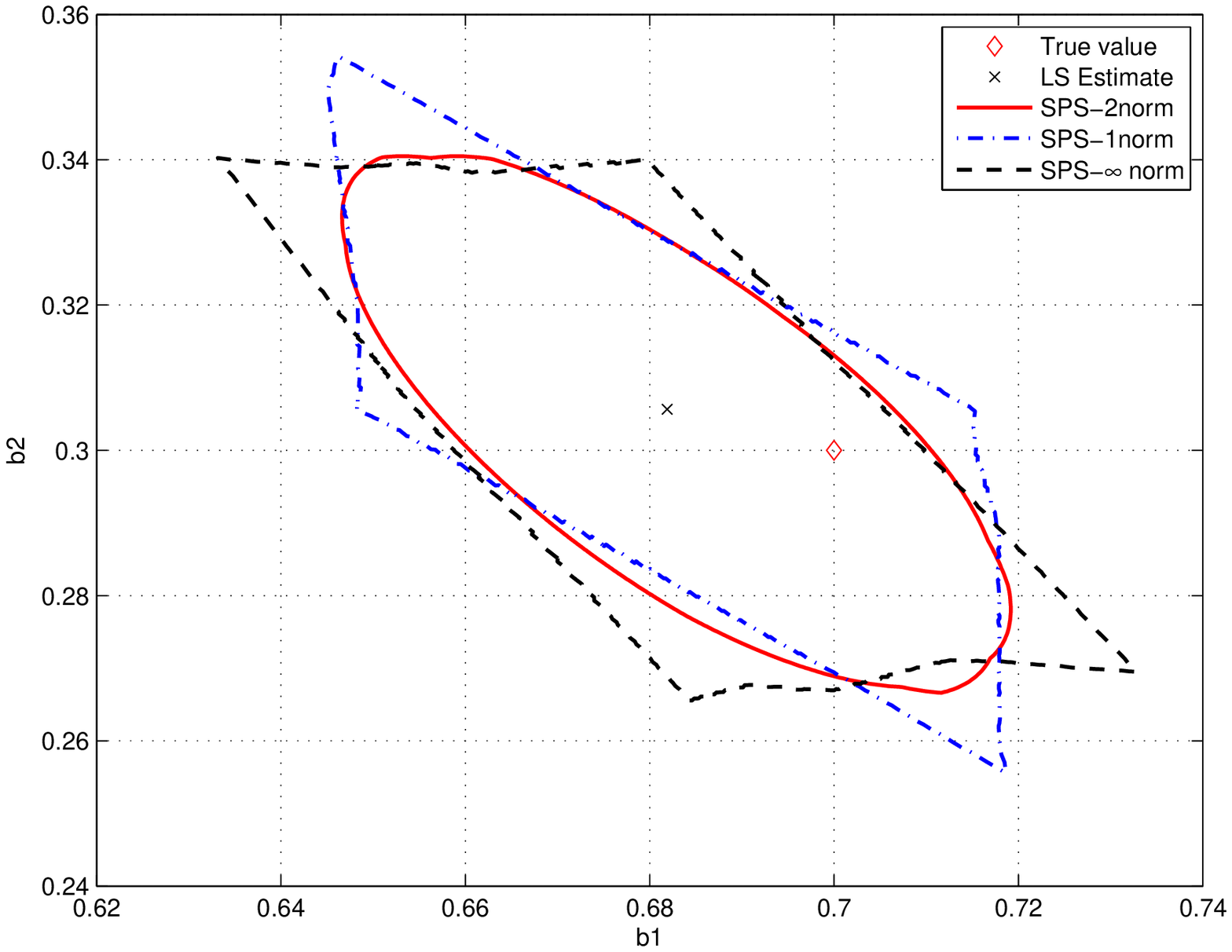}
\vspace{-5mm}
\end{center}\caption{$95$\% confidence regions using various norms, $n=400$, $m=100$.}\label{N400b}
\end{figure}

The confidence region is constructed as the values of $\theta$ for which at least 5 of the $||S_i(\theta)||^2$, $i=1,\ldots,99$, functions are larger than $||S_0(\theta)||^2$. Here $m=100$ and $q=5$ and it follows from Theorem \ref{theorem-exact-prob} that the constructed region contains the true parameter with exact probability $1-\frac{5}{100} = 95\%$.

The SPS confidence region is shown in Figure \ref{N25} together with the outer approximation and the confidence region based on the asymptotic system identification theory. 

It can be observed that the non-asymptotic SPS region is similar in shape to, and not too different in size from, the asymptotic confidence region, while it has the advantage that it is guaranteed to contain the true parameter with exact probability $95\%$, unlike the ones based on asymptotic results.

Next, the number of data points were increased to $n=400$, still with $m=100$ and $q=5$, and the confidence regions in Figure \ref{N400} were obtained. As can be seen, the differences between the various regions get smaller, and the SPS confidence region concentrates around the true parameter as $n$ increases.

Finally, the effect of using $\| \cdot\|_1$ or $\| \cdot\|_{\infty}$ norms, instead of $\| \cdot\|_2$, was considered. Recalling that the SPS regions are exact for each norm, the theory can be used to establish a precise confidence for each construction. The obtained confidence regions are illustrated in Figure \ref{N400b}. 

\subsection{Choice of $m$ and $q$}
The probability of the SPS confidence region is $1-q/m$ and hence there are many choices of $q$ and $m$ that give the same probability. Based on experience, selecting $q$ and $m$  too low increases the stochastic volatility in the SPS construction, and, consequently, the average area of the confidence regions tends to be larger. However, a saturation effect occurs so that pushing $q$ and $m$ beyond certain values has no practical benefit. This is illustrated in Table \ref{qmtable} where the average area of the 95\% confidence regions  based on 500 simulations of the same system as in the previous section with $n=25$ is evaluated. 
As we can see, the average area decreases as $m$ increases, but there is  little reduction in the average area by increasing $m$ beyond 200.

{\renewcommand{\arraystretch}{1.3}
\begin{table}
\vspace{1mm}
\begin{center}
\begin{tabular}{|l||r|r|r|r|r|r|}\hline
$m$ & $20$& $60$&  $100$ &  $200$ &$400$ &  $600$\\ \hline\hline
average area &    0.1041   & 0.0837  &  0.0806  &  0.0788  &  0.0778  &  0.0777\\ \hline
\end{tabular}
\end{center}
\caption{Average area, $n=25$.}
\label{qmtable}
\vspace*{-5mm}
\end{table}
}

In all simulation examples above Laplacian noise was used which is heavy-tailed. However, very similar results were obtained with, for example, uniform and Gaussian noises
\cite{WCC2013}.

\subsection{Comparing with Exact Confidence Ellipsoids Based on the F-distribution}

In the special case that the noise $\{N_t\}$ is a sequence of i.i.d. Gaussian random variables, the quantity
\[
\frac{n}{d} \frac{1}{\hat{\sigma}_n^2}(\theta^* - \hat{\theta}_n)^\mathrm{T}R_n(\theta^* - \hat{\theta}_n)
\]
is distributed according to an $F(d,n-d)$ distribution where $d$ is the number of parameters in $\theta$ and $\hat{\sigma}^2_n$ is given by (\ref{noise-variance-estimate}).

In this case
\[
\widetilde{\Theta}_n\, \triangleq \, \bigg\{\, \theta \in \mathbb{R}^d\, :\, (\theta - {\hat{\theta}_n})^\mathrm{T}\, R_n\, (\theta - {\hat{\theta}_n}) \, \leq \, \frac{\mu d \hat{\sigma}^2_n}{n} \,\bigg\}
\]
contains the true parameter value with exact probability $F_F(\mu)$ where  $F_F(\mu)$ is the cumulative distribution of an $F(d,n-d)$ distributed random variable.

SPS constructs exact confidence regions under much weaker conditions, and even when the noise is i.i.d. Gaussian we do not loose much since the SPS confidence regions are comparable to those obtained using the $F$-distribution as the following results show.

We consider the same system as in the previous section with $n=25$ data points. This time $\{N_t\}$ was a sequence of i.i.d. zero mean Gaussian random variables with variance 0.1. Figure \ref{SPSFdist1} shows the 95\% confidence regions we obtained in four simulation trials, and similar plots with $n=200$ are shown in Figure \ref{SPSFdist2}. Table \ref{SPSFdisttable} gives the average area of the confidence ellipsoids based on 1000 Monte Carlo simulations with $n=25$ and $n=200$. The average area increases by 20\% ($n=25$) and 6\% ($n=200$), when the prior information about the noise is reduced from i.i.d. Gaussian to independent and symmetrically distributed.

\begin{figure}[t]
\vspace*{-2mm}
\setlength{\epsfysize}{8.6cm} \setlength{\epsfxsize}{8.8cm}
\begin{center}
\leavevmode \epsfbox{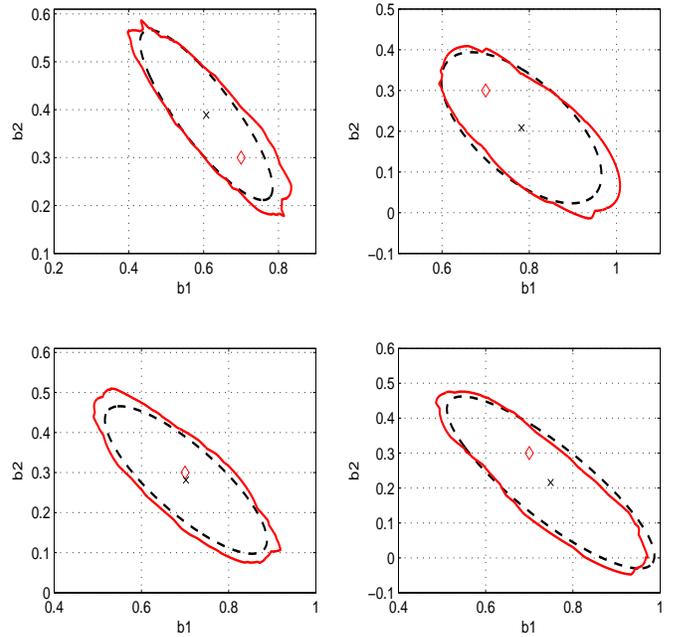}
\end{center}
\vspace{-4mm}
\caption{$95$\% confidence regions, $n=25$, $m=100$. The solid line gives the SPS region. The dashed line gives the confidence ellipsoid based on the $F$-distribution.}\label{SPSFdist1}
\end{figure}
\begin{figure}
\setlength{\epsfysize}{8.6cm} \setlength{\epsfxsize}{8.8cm}
\begin{center}
\leavevmode \epsfbox{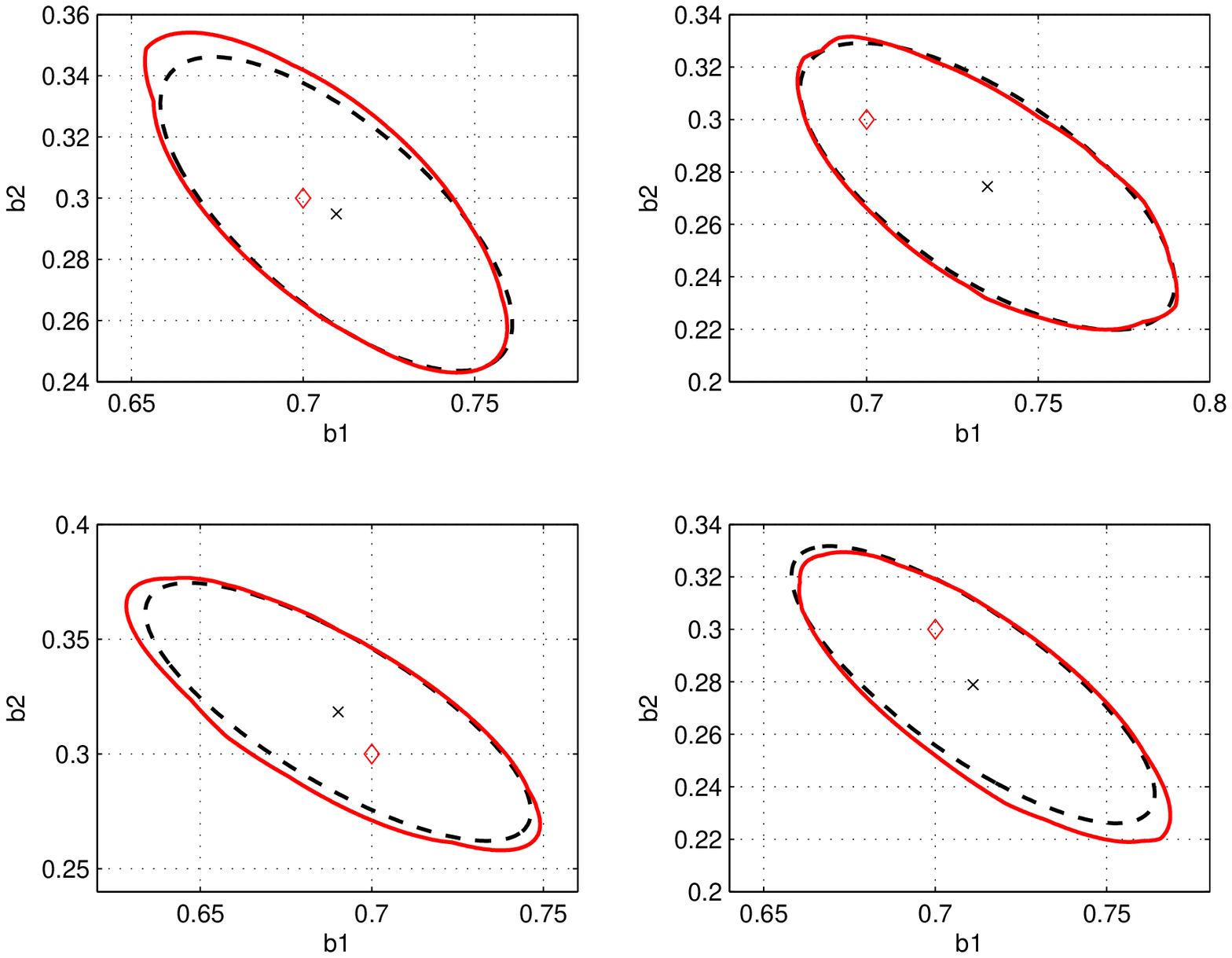}
\end{center}
\vspace{-4mm}
\caption{$95$\% confidence regions, $n=200$, $m=100$. The solid line gives the SPS region. The dashed line gives the confidence ellipsoid based on the $F$-distribution.}\label{SPSFdist2}
\end{figure}
\begin{figure}
\medskip
\setlength{\epsfysize}{6.5cm} \setlength{\epsfxsize}{8.0cm}
\begin{center}
\leavevmode \epsfbox{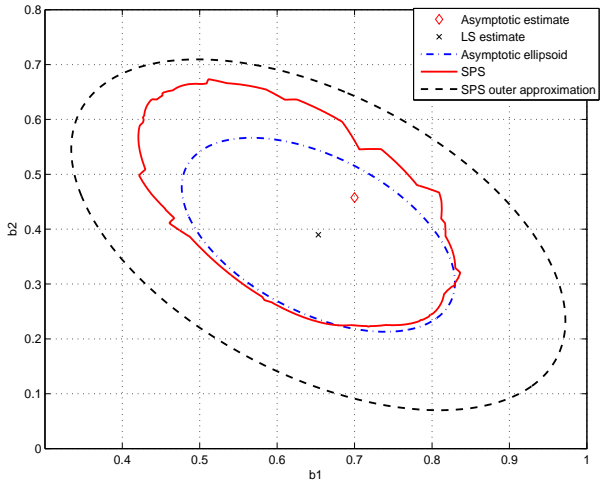}
\vspace{-4mm}
\end{center}\caption{$95$\% confidence regions, $n=25$. The true system is a third order system, while the model is second order.}\label{undermodfig}
\medskip
\end{figure}
\begin{figure}[t]
\setlength{\epsfysize}{8.6cm} \setlength{\epsfxsize}{8.8cm}
\begin{center}
\leavevmode \epsfbox{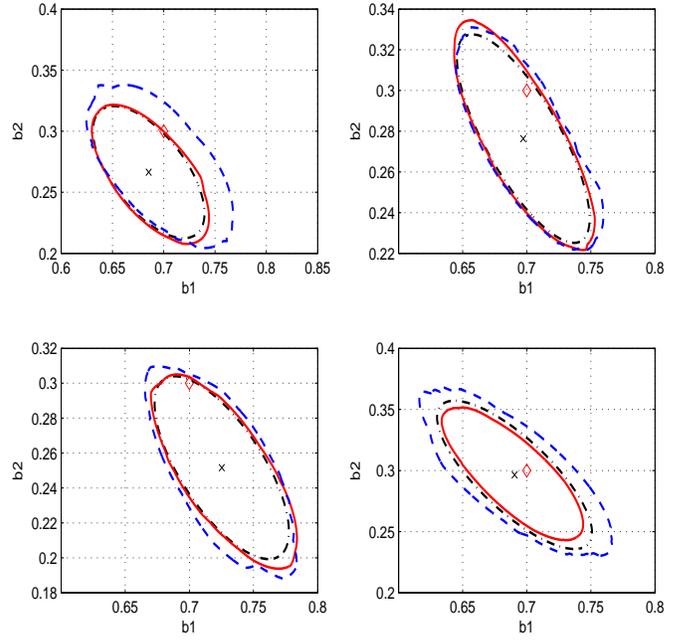}
\end{center}
\vspace{-4mm}
\caption{$95$\% confidence regions, $n=200$, $m=100$. The solid line gives the standard SPS region and the dashed  line gives the block SPS region. The dash dotted line gives the confidence ellipsoid based on asymptotic system identification theory.}\label{chunkspsconfidencefig}
\end{figure}
\medskip

{\renewcommand{\arraystretch}{1.3}
\begin{table}
\vspace{3mm}
\begin{center}
\begin{tabular}{|l||r|r|r|}\hline
 & SPS& $F$-distribution \\ \hline\hline
$n=25$&   0.07876  &  0.065658  \\ \hline
$n=200$&   0.00689  &  0.00650  \\ \hline
\end{tabular}
\end{center}
\caption{Average area.}
\label{SPSFdisttable}
\vspace{-4mm}
\end{table}
}

\subsection{Undermodelling}
The true data generating system is now given by
\[
  Y_t=b^*_1U_{t-1}+b^*_2U_{t-2}+b^*_3U_{t-3}+N_t,
\]
where $b^*_1=0.7$,  $b^*_2=0.3$ and $b^*_3=0.21$ are the true system parameters. $N_t$ and $U_t$ are as in Section \ref{secfir}.

The model class is still the class of all second order FIR systems with predictors
\[
 \hat{Y}_t(\theta)=b_1U_{t-1}+b_2U_{t-2}.
\]
In this case the model class is not rich enough to contain the true system. The asymptotic least squares estimate as the number of data points tends to infinity is the value of $\theta=[b_1, b_2]^\mathrm{T}$ such that $E(Y_t-\hat{Y}_t(\theta))^2$ is minimised.  For the  input signal used
these values are $\hat{b}^*_1=b_1^*=0.7$ and $\hat{b}^*_2=b_2^*+0.75b^*_3=0.4575$. A Monte Carlo simulation with {1\,000\,000} run found that the 95\% confidence region contained the asymptotic LS estimate with empirical probability 0.9509 with $n=25, m=100$ and $q=5$, which shows that in this example SPS exhibits robustness with respect to undermodelling. 
A typical result is shown in Figure \ref{undermodfig}.

\subsection{Assumptions on the Noise are not Satisfied}\label{blockSPSsim}
In this example, the assumptions on the noise are not satisfied. The system is the same as in the previous sections, but the noise is now the autoregressive process
\[ N_t=0.3N_{t-1}+\sqrt{1-0.3^2}W_t,
\]
where $\{W_t\}$ is i.i.d. Gaussian with variance 0.1. $n=200$ data points are available and the aim is as before to generate a 95\% confidence region. Due to the correlation in the noise both standard SPS and asymptotic system identification theory fail to produce confidence regions with the required probability. However, by using block SPS as described in Section \ref{blockSPS} where the random signs kept their values 1 or -1 for 10 consecutive values we got a confidence probability much closer to the desired 95\% than by using standard SPS or asymptotic system identification theory as shown in Table \ref{empprobARnoise}.

Examples of confidence regions obtained in four simulation trials are shown in Figure  \ref{chunkspsconfidencefig}. This demonstrates that block SPS works well also when the assumptions on the noise are not satisfied, and although the results are not precisely guaranteed anymore, it still gives good approximations.

Further, we see that we do not loose too much by using block SPS instead of the standard SPS when the assumptions on the noise are satisfied. Table \ref{chunkspsarea} shows the average area of the 95\% confidence sets based on 1000 Monte Carlo simulations  when the noise was i.i.d. Gausian with variance 0.1.  The average area only increased with  8.9\%.

{\renewcommand{\arraystretch}{1.3}
\begin{table}[t]
\begin{center}
\begin{tabular}{|r|r|r|}\hline
  SPS& Block SPS & Asymptotic theory \\ \hline\hline
      0.888  &  0.944  &     0.883
 \\ \hline
\end{tabular}
\end{center}
\caption{Empirical probabilities based on $10^6$ Monte Carlo simulations.}
\label{empprobARnoise}
\end{table}
}
{\renewcommand{\arraystretch}{1.3}
\begin{table}[h]
\begin{center}
\begin{tabular}{|l|r|r|r|}\hline
  SPS& Block SPS \\ \hline\hline
  0.00682  &  0.00743  \\ \hline
\end{tabular}
\end{center}
\caption{Average area, $n=200$.}
\label{chunkspsarea}
\vspace*{-4mm}
\end{table}
}
{\renewcommand{\arraystretch}{1.3}
\begin{table}[h]
\begin{center}\begin{tabular}{|l||r|}\hline
 & Relative increase per dimension\\ 
 Data points& of the ellipsoidal over-bound\\ \hline\hline
$n=200$ &   1.78
\\ \hline
$n=800$ &   1.34
\\ \hline
$n=3200$ & 1.17
\\ \hline
\end{tabular}
\end{center}
\caption{Volumes. 8th order system.}
\label{areashighord}
\vspace*{-4mm}
\end{table}
}

\subsection{Higher Order System}
In this experiment, the data generating system is an eight order FIR system, that is,
\begin{eqnarray*}
Y_t&=&b^*_1\,U_{t-1}+b^*_2\,U_{t-2}+b^*_3\,U_{t-3}+b^*_4\,U_{t-4}+b^*_5\,U_{t-5} \\ && + b^*_6\,U_{t-6} + b^*_7\,U_{t-7} + b^*_8\,U_{t-8}+N_t,
\end{eqnarray*}
with $\theta^*=[\,0.7, 0.3, 0.21, 0.2, 0.15, 0.25, 0.1, 0.05\,]^\mathrm{T}$. Processes $\{U_t\}$ and $\{N_t\}$ are the same as in Section \ref{secfir}.

We ran $1000$ Monte Carlo simulations with $n=200,\ 800$ and $3200$, and computed the ellipsoidal over-bound for the $95$\% confidence region using $m=100$ and $q=5$.

The computational time for computing a single ellipsoidal over-bound with 3200 data points was around $23$ seconds on a standard laptop using Yalmip and SDPT3, showing that the computational burden of
the approximation is quite modest.

Table \ref{areashighord} gives the relative increase per dimension of the ellipsoidal over-bound
as compared with the ellipsoid of the asymptotic theory.
The confidence ellipsoids based on asymptotic theory are smaller than the ones based on the ellipsoidal over-bound, but the difference gets smaller as $n$ increases, and the probability is guaranteed using SPS while it is not when using asymptotic theory.

\section{Summary and Conclusion}
\label{conclusions}
In this paper a new system identification method called {\em Sign-Perturbed Sums} (SPS) has been introduced. SPS allows the construction of guaranteed non-asymptotic confidence regions, which are built around the least-squares estimate and contain the true system parameter with a user-chosen exact probability for any finite data set. SPS works under mild statistical assumptions on the system noise, and it is {\em non-conservative}, i.e., its confidence probability is exact. In addition, it was shown that the SPS confidence regions are 
{\em star convex} with the LS estimate as a star center. 

Evaluating whether a given parameter value $\theta$ belongs to the SPS confidence region is a task that can be carried out at low computational cost. This makes SPS an effective method to apply when only a finite number of candidate $\theta$ values have to be tested. On the other hand, finding the precise boundary of the SPS set can be computationally demanding in general. In order to overcome this issue, an algorithm has been introduced that provides an outer approximation of the SPS region in the form of an ellipsoid. It was demonstrated that such over-bound can be efficiently computed by convex programming methods.

Simulation experiments demonstrated that the SPS method works well, and that the confidence regions have similar size and shape as the heuristic ellipsoids of the asymptotic theory or the exact ellipsoids based on the $F$-distribution when the noise is i.i.d. Gaussian.

In this paper we assumed that the regressors are deterministic. While it is easy to generalize our results to the case of random regressors that are independent of the noise, the extension to the case where the regressors can depend on the noise terms is non-trivial. Generalizing the method to that case is of high practical importance. We leave this
to further work, noting that some preliminary results addressing this issue were presented in earlier conference papers \cite{Csaji2012a,Csaji2012b}.

\bibliographystyle{IEEEtran}
\bibliography{SPS-Journal-IEEE}

\begin{thebibliography}{10}
\providecommand{\url}[1]{#1}
\csname url@samestyle\endcsname
\providecommand{\newblock}{\relax}
\providecommand{\bibinfo}[2]{#2}
\providecommand{\BIBentrySTDinterwordspacing}{\spaceskip=0pt\relax}
\providecommand{\BIBentryALTinterwordstretchfactor}{4}
\providecommand{\BIBentryALTinterwordspacing}{\spaceskip=\fontdimen2\font plus
\BIBentryALTinterwordstretchfactor\fontdimen3\font minus
  \fontdimen4\font\relax}
\providecommand{\BIBforeignlanguage}[2]{{%
\expandafter\ifx\csname l@#1\endcsname\relax
\typeout{** WARNING: IEEEtran.bst: No hyphenation pattern has been}%
\typeout{** loaded for the language `#1'. Using the pattern for}%
\typeout{** the default language instead.}%
\else
\language=\csname l@#1\endcsname
\fi
#2}}
\providecommand{\BIBdecl}{\relax}
\BIBdecl

\bibitem{Ljung1999}
L.~Ljung, \emph{System Identification: Theory for the User}, 2nd~ed.\hskip 1em
  plus 0.5em minus 0.4em\relax Prentice-Hall, Upper Saddle River, 1999.

\bibitem{Ljung1994}
L.~Ljung and T.~Glad, \emph{Modeling of Dynamic Systems}.\hskip 1em plus 0.5em
  minus 0.4em\relax Prentice Hall, 1994.

\bibitem{Ljung2010}
L.~Ljung, ``Perspectives on system identification,'' \emph{Annual Reviews in
  Control}, vol.~34, no.~1, pp. 1--12, 2010.

\bibitem{Soderstrom1989}
T.~S{\"o}derstr{\"o}m and P.~Stoica, \emph{System Identification}.\hskip 1em
  plus 0.5em minus 0.4em\relax Prentice Hall International, Hertfordshire, UK,
  1989.

\bibitem{Gevers2006}
M.~Gevers, ``A personal view of the development of system identification: a
  30-year journey through an exciting field,'' \emph{Control Systems, IEEE},
  vol.~26, no.~6, pp. 93--105, 2006.

\bibitem{Campi2005}
M.~C. Campi and E.~Weyer, ``Guaranteed non-asymptotic confidence regions in
  system identification,'' \emph{Automatica}, vol.~41, pp. 1751--1764, 2005.

\bibitem{Campi2010}
------, ``Non-asymptotic confidence sets for the parameters of linear transfer
  functions,'' \emph{IEEE Transasctions on Automatic Control}, vol.~55, pp.
  2708--2720, 2010.

\bibitem{Campi2009}
M.~C. Campi, S.~Ko, and E.~Weyer, ``Non-asymptotic confidence regions for model
  parameters in the presence of unmodelled dynamics,'' \emph{Automatica},
  vol.~45, pp. 2175--2186, 2009.

\bibitem{Dalai2007}
M.~Dalai, E.~Weyer, and M.~C. Campi, ``Parameter identification for non-linear
  systems:\,guaranteed confidence regions through {LSCR},'' \emph{Automatica},
  vol.~43, pp. 1418--1425, 2007.

\bibitem{Weyer2009}
E.~Weyer, S.~Ko, and M.~C. Campi, ``Finite sample properties of system
  identification with quantized output data,'' in \emph{Proceedings of the 48th
  IEEE Conference on Decision and Control}, 2009, pp. 1532--1537.

\bibitem{Weyer2009b}
------, ``A randomised subsampling method for change detection,'' in
  \emph{Proceedings of the SafeProcess'09 Conference, Barcelona, Spain}, 2009.

\bibitem{Schoukens2013}
J.~Schoukens, Y.~Rolain, G.~Vandersteen, and R.~Pintelon, ``Study of small data
  set efficiency losses in system identification: The {FIR} case,'' in
  \emph{Proceedings of the 11th IFAC International Workshop on Adaptation and
  Learning in Control and Signal Processing}, 2013, pp. 68--73.

\bibitem{Dabbene2014}
F.~Dabbene, M.~Sznaier, and R.~Tempo, ``Probabilistic optimal estimation with
  uniformly distributed noise,'' \emph{IEEE Transactions on Automatic Control},
  vol.~59, no.~8, pp. 2113--2127, 2014.

\bibitem{Aguero2012}
J.~C. Aguero, C.~R. Rojas, H.~Hjalmarsson, and G.~C. Goodwin, ``Accuracy of
  linear multiple-input multiple-output ({MIMO}) models obtained by maximum
  likelihood estimation,'' \emph{Automatica}, vol.~48, pp. 632--637, 2012.

\bibitem{Granichin2012}
O.~N. Granichin, ``The nonasymptotic confidence set for parameters of a linear
  control object under an arbitrary external disturbance,'' \emph{Automation
  and Remote Control}, vol.~73, no.~1, pp. 20--30, 2012.

\bibitem{DenDekker2008}
A.~J. {den Dekker}, X.~Bombois, and P.~M.~J. {Van den Hof}, ``Finite sample
  confidence regions for parameters in prediction error identification using
  output error,'' in \emph{IFAC World Congress}, 2008, pp. 5024--5029.

\bibitem{Hjalmarsson2006}
H.~Hjalmarsson and B.~Ninness, ``Least-squares estimation of a class of
  frequency functions: A finite sample variance expression,''
  \emph{Automatica}, vol.~42, pp. 589--600, 2006.

\bibitem{Csaji2012a}
B.~\relax{Cs}.\ Cs\'aji, M.~C. Campi, and E.~Weyer, ``Non-asymptotic confidence
  regions for the least-squares estimate,'' in \emph{Proceedings of the 16th
  IFAC Symposium on System Identification}, 2012, pp. 227--232.

\bibitem{Csaji2012b}
------, ``A method for constructing exact finite-sample confidence regions for
  general linear systems,'' in \emph{Proceedings of the 51st IEEE Conference on
  Decision and Control}, 2012, pp. 7321--7326.

\bibitem{WCC2013}
E.~Weyer, B.~\relax{Cs}.\ Cs\'aji, and M.~C. Campi, ``Guaranteed non-asymptotic
  confidence ellipsoids for {FIR} systems,'' in \emph{Proceedings of the 52st
  IEEE Conference on Decision and Control}, 2013, pp. 7162--7167.

\bibitem{VanDenHof1995}
P.~M.~J. {Van den Hof}, P.~S.~C. Heuberger, and J.~Bokor, ``System
  identification with generalized orthonormal basis functions,''
  \emph{Automatica}, vol.~31, pp. 1821--1834, 1995.

\bibitem{VanDenHof2005}
P.~Van~den Hof and B.~Ninness, ``System identification with generalized
  orthonormal basis functions,'' in \emph{Modelling and Identification with
  Rational Orthogonal Basis Functions}.\hskip 1em plus 0.5em minus 0.4em\relax
  Springer London, 2005, pp. 61--102.

\bibitem{Efron1993}
B.~Efron and R.~Tibshirani, \emph{An Introduction to the Bootstrap}.\hskip 1em
  plus 0.5em minus 0.4em\relax Chapman \& Hall, New York, 1993.

\bibitem{davidson2006}
R.~Davidson and J.~G. MacKinnon, ``Bootstrap methods in econometrics,''
  \emph{Palgrave handbook of econometrics}, vol.~1, pp. 812--38, 2006.

\bibitem{Godfrey2009}
L.~Godfrey, \emph{Bootstrap Tests for Regression Models}.\hskip 1em plus 0.5em
  minus 0.4em\relax Palgrave Macmillan, 2009.

\bibitem{KiefferWalter2013}
M.~Kieffer and E.~Walter, ``Guaranteed characterization of exact non-asymptotic
  confidence regions as defined by {LSCR} and {SPS},'' \emph{Automatica},
  vol.~49, pp. 507--512, 2013.

\bibitem{KiefferWalter2013a}
------, ``Guaranteed characterization of exact non-asymptotic confidence
  regions in nonlinear parameter estimation nonlinear control systems,'' in
  \emph{Proceedings of the 9th IFAC Symposium on Nonlinear Control Systems},
  2013, pp. 56--61.

\bibitem{Boyd2009}
S.~Boyd and L.~Vandenberghe, \emph{Convex Optimization}.\hskip 1em plus 0.5em
  minus 0.4em\relax Cambridge University Press, 2009.

\bibitem{Lovberg2004}
L\"ofberg, ``Yalmip : A toolbox for modeling and optimization in {MATLAB},'' in
  \emph{Proceedings of the CACSD Conference,Taipei, Taiwan}, 2004.

\bibitem{Kallenberg2001}
O.~Kallenberg, \emph{Foundations of Modern Probability}, 2nd~ed.\hskip 1em plus
  0.5em minus 0.4em\relax Springer-Verlag, New York, 2001.

\bibitem{Zhang2005}
F.~Zhang, \emph{The Schur Complement and its Applications}.\hskip 1em plus
  0.5em minus 0.4em\relax Springer, 2005.

\bibitem{Zhang2011}
------, \emph{Matrix Theory: Basic Results and Techniques}.\hskip 1em plus
  0.5em minus 0.4em\relax Springer, 2011.

\end{thebibliography}

\appendices
\section{Proof of Theorem \ref{theorem-exact-prob}:Exact Confidence}
\label{appendix-exactness-proof}
We begin with a definition and some lemmas.

\medskip
\begin{definition}
\label{unif-order}
{\em Let $Z_1, \dots, Z_{k}$ be a finite collection of random variables and $\succ$ a strict total order. If for all permutations $i_1, \dots, i_{k}$ of indices $1,\dots, k$ we have
\begin{equation*}
\mathbb{P}(Z_{i_k} \succ Z_{i_{k-1}} \succ \dots \succ Z_{i_{1}}) = \frac{1}{k!},
\end{equation*}
then we call $\{Z_i\}$ uniformly ordered w.r.t.\ order $\succ$.}
\end{definition}

\medskip
\begin{lemma}
\label{lemma-random-signs} {\em Let $\alpha, \beta_1, \dots, \beta_k$ be i.i.d.\ random signs,
then the random variables $\alpha, \alpha \cdot \beta_1, \dots, \alpha \cdot \beta_k$ are i.i.d.\ random signs.}
\end{lemma}
\smallskip
{\em Proof.}
Let $c_0, c_1, \dots, c_k$ be a fixed vector of signs, i.e., $c_i \in \{-1, 1\}$. Then, we have
\begin{equation*}
\mathbb{P}(\alpha = c_0,\, \alpha \beta_1 = c_1,\, \dots,\, \alpha \beta_k = c_k)
\end{equation*}
\begin{equation*}
=\mathbb{P}(\alpha = c_0,\, \beta_1 = c_0 c_1,\, \dots,\, \beta_k = c_0 c_k)
\end{equation*}
\begin{equation*}
=\mathbb{P}(\alpha = c_0)\,\mathbb{P}(\beta_1 = c_0c_1) \dots \,\mathbb{P}(\beta_k = c_0 c_k)
\end{equation*}
\begin{equation*}
=\mathbb{P}(\alpha = c_0)\,\mathbb{P}(\alpha \beta_1 = c_1) \dots \,\mathbb{P}(\alpha\beta_k = c_k),
\end{equation*}
where we have used that the original collection was independent, and that $\alpha \beta_i$ and $\beta_i$ has the same probability mass function, i.e., for all signs $a, b\in \{-1, 1\}$: $\mathbb{P}(\beta_i = b) = \mathbb{P}(\alpha \beta_i = a) = 1/2$.
$\Box$

\medskip
\begin{lemma}
\label{lemma-realizations}
{\em Let $X$ and $Y$ be two independent, $\mathbb{R}^d$-valued and $\mathbb{R}^k$-valued random vectors, respectively. Let us consider a (measurable) function
$g: \mathbb{R}^d \times \mathbb{R}^k \to \mathbb{R}$ and a (measurable) set $A \subseteq \mathbb{R}$. If we have $\,\mathbb{P}(\,g(x,Y) \in A\,) = p$, for all (constant) $\,x \in \mathbb{R}^d$, then we also have $\,\mathbb{P}(\,g(X,Y) \in A\,) = p$.}
\end{lemma}
\smallskip
{\em Proof.}
Define $\mathbb{I}_A$ as follows
\begin{equation*}
\mathbb{I}_A(x, y) \,\triangleq\,
\left\{
\begin{array}{ll}
1 \hspace{3mm}& \mbox{if}\hspace{2mm} g(x, y) \in A,\\
0 \hspace{3mm}& \mbox{otherwise},\\
\end{array}
\right.
\end{equation*}
which is the indicator function of the event that $g(X, Y) \in A$. Now, let us define function $i_A:\mathbb{R}^d \to \mathbb{R}$ as
$i_A(x) \,\triangleq\, \mathbb{E}\left[\,\mathbb{I}_A(x, Y) \,\right]$,
where $x \in \mathbb{R}^d$ is a constant, therefore, $i_A(x)$ is a number (non-random). We know
that for all $x \in \mathbb{R}^d$ we have $i_A(x) = \mathbb{P}(\,g(x,Y) \in A\,) = p$. Then, by applying the properties of the conditional expectation \cite{Kallenberg2001}, we have that
\begin{equation*}
\mathbb{P}(\,g(X,Y) \in A\,) = \mathbb{E}\left[\,\mathbb{I}_A(X, Y) \,\right]
\end{equation*}
\begin{equation*}
= \mathbb{E}\left[\,\mathbb{E}\left[\,\mathbb{I}_A(X, Y)\,|\,X \,\right]\,\right] = \mathbb{E}\left[\,i_A(X)\,\right] = \mathbb{E}\left[\,p\,\right] = p,
\end{equation*}
which completes the proof of the lemma. $\Box$
\medskip

The following lemma highlights an important property of the $\succ_{\pi}$ relation that was introduced in Section \ref{sps-method}.
\medskip
\begin{lemma}
\label{lemma-iid-case}
{\em Let $Z_1, \dots, Z_{k}$ be real-valued, i.i.d. random variables. Then, they are uniformly ordered w.r.t.\ $\succ_{\pi}$.}
\end{lemma}
\smallskip
{\em Proof.} Since $\succ_{\pi}$ is a total order which resolves ties, there is a {\em unique} ordering for all realizations $Z_1(\omega), \dots, Z_k(\omega)$ and $\pi(\omega)$. Therefore, the events $Z_{i_k} \succ_{\pi} \dots \succ_{\pi} Z_{i_{1}}$ define a complete system of events.
There are $k!$ such orderings, thus, in order to complete the proof we need to show that each such ordering has the same probability. Let us select two orderings $Z_{i_k} \succ_{\pi} \dots \succ_{\pi} Z_{i_{1}}$ and $Z_{j_k} \succ_{\pi} \dots \succ_{\pi} Z_{j_{1}}$, which have probabilities, $p_i$ and $p_j$, respectively. We will show that $p_i = p_j$. First, we define some new random variables $Z'_{i_1} \triangleq Z_{j_{1}}, \dots, Z'_{i_k} \triangleq Z_{j_{k}}$. Then, $Z'_{i_k} \succ_{\pi} \dots \succ_{\pi} Z'_{i_{1}}$ has the same probability as $Z_{j_k} \succ_{\pi} \dots \succ_{\pi} Z_{j_{1}}$, since the corresponding variables are the same and $\pi$ is completely {\em symmetric} with respect to the indices. Because $\{Z_k\}$ are i.i.d.,
$Z'_1, \dots, Z'_{k}$ has the same distribution as $Z_1, \dots, Z_{k}$. Then, $Z'_{i_k} \succ_{\pi} \dots \succ_{\pi} Z'_{i_{1}}$ must have the same probability as $Z_{i_k} \succ_{\pi} \dots \succ_{\pi} Z_{i_{1}}$, namely $p_i$. Thus, $p_j = p_i$. $\Box$

\subsection*{Proof of Theorem \ref{theorem-exact-prob}}
By construction, parameter $\theta^*$ is in the confidence region if
$\mathcal{R}(\theta^*) \leq m-q$.
This means that  $\|S_0(\theta^*)\|^2$ takes one of the positions $1, \dots, m-q$ in the ascending order (w.r.t. $\succ_{\pi}$) of variables $\{\|S_i(\theta^*)\|^2\}$. We are going to prove that the $\{\|S_i(\theta^*)\|^2\}$ are {\em uniformly ordered}, hence $\|S_0(\theta^*)\|^2$ takes each position in the ordering with probability $1/m$, thus its rank is at most $m-q$ with probability $1-q/m$.

First, note that for $\theta=\theta^*$, {\em all} $S_i(\cdot)$ functions have the form
\begin{equation*}
S_i(\theta^*) = R_n^{-\frac{1}{2}} \frac{1}{n} \sum_{t=1}^{n}{\, \alpha_{i,t} \, \varphi_t N_t},
\end{equation*}
for all $i \in \{0, \dots, m-1\}$, where $\alpha_{0,t} \triangleq 1$, $t \in \{1, \dots, n\}$.

Therefore, all the $S_i(\cdot)$ functions depend on the perturbed noise sequence, $\{\alpha_{i,t}N_t\}$, via the {\em same} function for all $i$, which we denote by $S(\alpha_{i,1}N_{1}, \dots, \alpha_{i,n}N_n) \triangleq S_i(\theta^*)$.

Since each $N_t$ is symmetric, we know that $sign(N_t)$ and $|N_t|$ are independent. Then, for all $i$ and $t$, we introduce $\gamma_{i,t} \triangleq \alpha_{i,t}\, sign(N_t)$. Using that $\{\alpha_{i,t}\}$ are i.i.d. random signs, independent of the other random elements, $\{N_t\}$ are independent, and applying Lemma \ref{lemma-random-signs}, it follows that $\{\gamma_{i, t}\}$ are not only independent of $\{|N_t|\}$, but also i.i.d.\ random signs.

After fixing a {\em realization} of $\{|N_t|\}$, called $\{v_t\}$, we define the real-valued variables $\{Z_i\}$ by
\begin{equation*}
Z_i \,\triangleq\, \|S(\gamma_{i,1} v_1, \dots, \gamma_{i,n} v_n)\|^2\!\!.
\end{equation*}

We know that, if the same (measurable) function is applied to each element of an i.i.d.\ sample, then the result will also be i.i.d.. Therefore, the $\{Z_i\}$ are i.i.d. random variables. Consequently, Lemma \ref{lemma-iid-case} can be applied to show that $\{Z_i\}$ are in fact uniformly ordered with respect to relation $\succ_{\pi}$.

So far we have proved the theorem assuming that the absolute values of the noises are constant, namely, the uniform ordering property was achieved by fixing a realization of $\{|N_t|\}$. However, the probabilities obtained are {\em independent of the particular realization} of $\{|N_t|\}$, hence, Lemma \ref{lemma-realizations} can be applied to relax fixing the realization (i.e., in Lemma \ref{lemma-realizations}, $X$ plays the role of $\{|N_t|\}$ and $Y$ incorporates the other random variables), and obtain the unconditional uniform ordering property of $\{\|S_i(\theta^*)\|^2\}$, from which the theorem follows. $\Box$

\section{Proof of Theorem \ref{theorem-starset}: Star Convexity}
\label{appendix-starset-proof}

Let $\Phi \triangleq [\,\varphi_1, \dots, \varphi_n\,]^\mathrm{T}$, $D_i \triangleq 1/n \cdot Diag(\alpha_{i,1}, \dots, \alpha_{i,n})$, $i \in \{1, \dots, m-1\}$, where $Diag(\cdot)$ is the diagonal matrix with its arguments on the main diagonal, and $Q_i \triangleq \Phi^\mathrm{T}D_i \Phi$. Notice that $\| S_i(\theta) \|^2$  and $\| S_0(\theta) \|^2$ are quadratic forms with Hessian $Q_i R_n^{-1} Q_i$ and $R_n$, respectively. The following lemma relates these Hessians.
\medskip
\begin{lemma}
\label{lemma-ordering} {\em For all $i$, we have $R_n \succeq Q_i R_n^{-1} Q_i$ in the L\"owner partial ordering, i.e., $R_n - Q_i R_n^{-1} Q_i$ is positive semidefinite}.
\end{lemma}
\smallskip
{\em Proof.} Since $R_n \succ 0$, a Schur complement argument, \cite{Zhang2005}, shows that the lemma statement is equivalent to the positive semidefiniteness of the matrix
\[B_i \triangleq
\left[\begin{array}{cc} R_n & Q_i \\ Q_i & R_n\end{array}\right] = \left[\begin{array}{cc} \frac{1}{n} \Phi^\mathrm{T} \Phi & \Phi^\mathrm{T}D_i \Phi \\ \Phi^\mathrm{T}D_i \Phi & \frac{1}{n} \Phi^\mathrm{T} \Phi\end{array}\right].
\]
Matrix $B_i$ can be decomposed as
\[ B_i =
\left[\begin{array}{cc} \Phi^\mathrm{T} & 0 \\ 0 & \Phi^\mathrm{T}\end{array}\right]
\left[\begin{array}{cc} \frac{1}{n} I & D_i \\ D_i & \frac{1}{n} I \end{array}\right]
\left[\begin{array}{cc} \Phi & 0 \\ 0 & \Phi \end{array}\right],
\]
so that $B_i$ is positive semidefinite if and only if the middle matrix is \cite{Zhang2011}. Resorting again to a Schur complement argument, the middle matrix is positive semidefinite if $\frac{1}{n^2}I - D_i D_i$ is positive semidefinite, which is clearly true since this latter matrix is in fact the zero matrix. $\Box$\\

Introduce now the set
\[
\mathcal{E}_i \, \triangleq \, \{\, \theta \in \mathbb{R}^d: \| S_0(\theta) \|^2 \leq \| S_i(\theta) \|^2 \,\}.
\]
Since in Lemma \ref{lemma-ordering} we have proven that the Hessian of $\| S_0(\theta) \|^2$ is no smaller than that of $\| S_i(\theta) \|^2$, it follows that $\| S_0(\theta) \|^2 - \| S_i(\theta) \|^2$ is a convex function, and $\{\, \theta \in \mathbb{R}^d: \| S_0(\theta) \|^2 - \| S_i(\theta) \|^2 \leq 0\,\} = \mathcal{E}_i$ is a convex set. Moreover, $\hat{\theta}_n \in \mathcal{E}_i$ since $\| S_0(\hat{\theta}_n) \|^2 = 0$. Likewise, one can prove that set
\[
\bar{\mathcal{E}}_i \, \triangleq \, \{\, \theta \in \mathbb{R}^d: \| S_0(\theta) \|^2 < \| S_i(\theta) \|^2 \,\}
\]
is convex or empty. Moreover, when it is not empty, it certainly contains $\hat{\theta}_n$. Indeed, $\| S_0(\hat{\theta}_n) \|^2 = 0$, so that for $\hat{\theta}_n$ not to be in $\bar{\mathcal{E}}_i$ it must be that $\| S_i(\hat{\theta}_n) \|^2 = 0$ too. In addition, since
$\| S_0(\theta) \|^2$ and $\| S_i(\theta) \|^2$
are quadratic forms, also their derivatives in $\hat{\theta}_n$ must be zero. Now, if by contradiction we assume that $\bar{\mathcal{E}}_i \neq \emptyset$, then there is a point $\bar{\theta} \in \bar{\mathcal{E}}_i$ and it holds that $\| S_0(\bar{\theta}) \|^2 < \| S_i(\bar{\theta}) \|^2$. Over the line segment connecting $\hat{\theta}_n$ to $\bar{\theta}$, function $\| S_i(\theta) \|^2 - \| S_0(\theta) \|^2$ then grows from $0$ to a positive value, and since it moves away from $\hat{\theta}_n$ with zero slope, it must have in some point in between $\hat{\theta}_n$ to $\bar{\theta}$ a positive curvature, a fact that contradicts Lemma \ref{lemma-ordering}.

We are now ready to establish the result in Theorem \ref{theorem-starset}.

Let
\[
\mathcal{E}_i^{\pi} \, \triangleq \, \{\, \theta \in \mathbb{R}^d : \| S_0(\theta) \|^2 \prec_{\pi} \| S_i(\theta) \|^2 \,\}.
\]
Notice that, depending on $\pi$, the set $\mathcal{E}^{\pi}_i$ is either $\mathcal{E}_i$ or $\bar{\mathcal{E}}_i$. Therefore, $\mathcal{E}_i^{\pi}$ is either empty or it is a convex set containing $\hat{\theta}_n$. Next,
note that the SPS confidence region $\widehat{\Theta}_n$ can be written as
\begin{equation*}
\widehat{\Theta}_n \,=\, \bigcup_{\substack{\mathcal{I} \subseteq \mathcal{M}\\ |\mathcal{I}|=q}} \,\bigcap_{i \in \mathcal{I}} \,\,\mathcal{E}^{\pi}_i,
\end{equation*}
where $\mathcal{M} = \{1, \dots, m-1\}$ and $|\cdot|$ denotes cardinality. To prove this note that if we take a finite index set of size $q$, $\{i_1, \dots, i_q\}$, then the intersection of the sets $\mathcal{E}^{\pi}_{i_j}$, $j \in \{1, \dots, q\}$, contains all the parameter values $\theta$ for which $\| S_0(\theta) \|^2$ is less than (w.r.t.\ $\prec_{\pi}$) all the functions $\| S_i(\theta) \|^2$ that have indexes from the index set $\{i_1, \dots, i_q\}$. When union is taken over all possible index sets, one obtains the set of parameter values $\theta$ for which $\| S_0(\theta) \|^2$ is less than (w.r.t.\ $\prec_{\pi}$) at least $q$ other functions $\| S_i(\theta) \|^2$, which is exactly the definition of the SPS confidence region. Finally, the theorem claim follows since unions and intersections of star convex sets having a common star center are themselves star convex with the same center. $\Box$

\end{document}